\documentclass[12pt,a4paper,hyper]{JHEP3}

\usepackage[utf8]{inputenc}
\usepackage{amsmath,epsfig}
\usepackage{amssymb,amsfonts}
\usepackage{graphicx}
\usepackage{slashed}
\title{\textsc{Quantum black holes, localization, 
 and 
 the topological string}}

\preprint{}
\author{
Atish Dabholkar$^{1, 2}$, Jo\~ao Gomes$^{1}$ and Sameer Murthy$^{1, 2}$\\

\it $^1${Laboratoire de Physique Th\'eorique et Hautes Energies (LPTHE)\\
\it{Universit\'e Pierre et Marie Curie-Paris 6; CNRS UMR 7589}\\
\it{Tour 24-25, 5$^{\grave{e}me}$ \'etage, Boite 126, 4 Place Jussieu} \\
\it {75252 Paris Cedex 05, France}}\\

\it $^2$Department of Theoretical Physics\\
\it Tata Institute of Fundamental Research\\
\it Homi Bhabha Rd, Mumbai 400 005, India\\

}

\abstract{We use localization to evaluate  the functional integral of string field theory   on $AdS_{2} \times S^{2}$ background  corresponding to the near horizon geometry of supersymmetric black holes in   4d compactifications with $\CN=2$ supersymmetry.  In particular, for a theory containing  $n_{v} + 1$ vector multiplets, we show that the functional integral localizes exactly onto an ordinary integral over a finite-dimensional submanifold in the field space labeling a continuous family of  instanton solutions in which auxiliary fields in the vector multiplets are excited with nontrivial dependence on $AdS_{2}$ coordinates. These localizing solutions  are \textit{universal} in that they follow from the off-shell supersymmetry transformations and do not depend on the choice of the action.  They are  parametrized by $n_{v} +1$ real parameters $\{ C^{I}\, ; I= 0, \ldots, n_{v} \}$ that correspond to 
the values of the auxiliary fields at the center of $AdS_{2}$.   In the Type-IIA frame, assuming D-terms evaluate to zero on the solutions for reasons of supersymmetry, the classical part of the integrand  equals the  absolute square of the partition function of the topological string as conjectured by Ooguri, Strominger, and Vafa; however evaluated at the off-shell values of scalar fields at the center of $AdS_{2}$.  In  addition, there are contributions from one-loop determinants, brane-instantons, and nonperturbative orbifolds that are in principle computable. These results thus provide a concrete method to compute exact quantum entropy  of these black holes  including all perturbative and nonperturbative corrections and can be used  to establish a  precise relation between the quantum degeneracies of black holes and the  topological string. }


\newenvironment{myenumerate}{
\begin{enumerate}
   \setlength{\itemsep}{1pt}
   \setlength{\parskip}{0pt}
   \setlength{\parsep}{0pt}}{\end{enumerate}}
\newenvironment{myitemize}{
\begin{itemize}
   \setlength{\itemsep}{1pt}
   \setlength{\parskip}{0pt}
   \setlength{\parsep}{0pt}}{\end{itemize}}

\newcommand{\IR}{\mathbb{R}}
\newcommand{\IZ}{\mathbb{Z}}

\def\slash#1{\rlap{\hbox{$\mskip 1 mu /$}}#1}      
\def\Slash#1{\rlap{\hbox{$\mskip 3 mu /$}}#1}      

\def\a{\alpha}

\def\m{\mu}

\def\e{\epsilon}
\def\w{\omega}
\def\h{\eta}

\def\O{{\Omega}}

\def\CF{{\cal F}}
\def\CS{{\cal S}}
\def\CL{{\cal L}}

\def\CN{{\cal N}}
\def\CO{{\cal O}}

\def\half{{\frac12}}

\def\IC{{\mathbb C}}

\def\CN{{\cal N}}

\def\bea{\begin{eqnarray}}
\def\eea{\end{eqnarray}}
\def\be{\begin{equation}}
\def\ee{\end{equation}}
\def\ba{\begin{align}}
\def\ea{\end{align}}
\def\bse{\begin{subequations}}
\def\ese{\end{subequations}}
\def\1F1{{}_1\!F_1}
\def\2F0{{}_2\!F_0}

\def\ve{\varepsilon}

\def\a{\alpha}

\def\h3{$\textrm{H}_3^+$}

\def\IC{{\mathbb C}}
\def\IR{{\mathbb R}}
\def\IZ{{\mathbb Z}}

\newcommand{\beq}{\begin{equation}}
\newcommand{\eeq}{\end{equation}}
\newcommand{\ber}{\begin{eqnarray}}
\newcommand{\eer}{\end{eqnarray}}

\def\be{\begin{eqnarray}}
\def\ee{\end{eqnarray}}

\newcommand{\cO}{{\cal O}}

\def\p{\partial}
\def\wh{\widehat}

\def\CN{{\cal N}}

\def\CF{{\cal F}}

\def\CL{{\cal L}}
\def\CV{{\cal V}}
\def\CO{{\cal O}}

\def\CE{{\cal E}}

\def\CS{{\cal S}}

\def\CE{{\cal E }}
\def\CV{{\cal V }}

\def\CS{{\cal S }}

\font\manual=manfnt
\def\dbend{\lower3.5pt\hbox{\manual\char127}}

\def\bar{\overline}
\def\CS{{\cal S}}
\def\CN{{\cal N}}

\def\rt2{\sqrt{2}}
\def\irt2{{1\over\sqrt{2}}}

\def\wt{\widetilde}
\def\ndt{\noindent}

\def\a{\alpha}
\def\w{\omega}

\font\cmss=cmss10
\font\cmsss=cmss10 at 7pt
\def\IL{\relax{\rm I\kern-.18em L}}
\def\IH{\relax{\rm I\kern-.18em H}}
\def\rlx{\relax\leavevmode}
\def\ZZ{\rlx\leavevmode\ifmmode\mathchoice{\hbox{\cmss Z\kern-.4em Z}}
 {\hbox{\cmss Z\kern-.4em Z}}{\lower.9pt\hbox{\cmsss Z\kern-.36em Z}}
 {\lower1.2pt\hbox{\cmsss Z\kern-.36em Z}}\else{\cmss Z\kern-.4em
 Z}\fi}

\begin{document}


\section{Introduction and summary of results}

The entropy of a black hole remains one of the most precious and precise clues about the microscopic structure of 
quantum gravity. In a consistent quantum theory of gravity, 
the black hole entropy must have a statistical interpretation in terms of underlying microstates in the quantum Hilbert 
space of the theory.  This is  a very strong constraint on the underlying short distance degrees of freedom, since it 
must hold in  \textit{any} phase of the theory that admits black holes.  For a theory under construction such as string 
theory, this requirement is a particularly useful guide especially since  we do not yet know which phase or `vacuum' of
the theory may correspond to the real world. In such a situation,  a profitable strategy is to focus on aspects of the
theory that must be universally true in all phases of the theory\footnote{The entropy itself will of course depend in an 
interesting way both on the phase and the states under consideration and is not expected to be universal.  What is  
universal is the constraint that this entropy must have a statistical interpretation in terms of the 
independently computable quantum degeneracies.}. 

By now, there is a very good statistical understanding of the entropy of  a  large class of supersymmetric charged 
black holes in several compactifications of string theory,  in the thermodynamic  limit of  large horizon area or large 
charges. In this limit,  the leading Bekenstein-Hawking entropy \cite{Bekenstein:1973ur, Hawking:1974sw} given 
by one quarter of the horizon area in Planck units precisely matches the logarithm of the degeneracy of the 
corresponding quantum microstates, as in the work of Strominger and Vafa \cite{Strominger:1996sh}. Given this 
beautiful agreement to leading order, it is important to figure out what exact formula this is an 
approximation to, and how one might systematically compute corrections to the leading answer for finite charges or 
equivalently for finite horizon area.  In the full quantum theory, both the macroscopic entropy and the microscopic 
degeneracy are expected to receive subleading corrections  at the perturbative as well as nonperturbative level. 

Finite size corrections to the leading  Bekenstein-Hawking entropy are physically very interesting for the following 
reason. The Bekenstein-Hawking entropy is in a sense a bit too universal in that it is  always given by a quarter of the 
horizon area. This is a consequence of the fact that it follows directly from the Einstein-Hilbert  action which has a 
universal form in all theories of gravity at very long distances. Finite size corrections, on the other hand,  arise from 
higher derivative corrections which are different in different phases. This dependence on the phase 
can yield  useful information about  different aspects of the short-distance theory. It is analogous to how one can learn 
about the fundamental microscopic Hamiltonian of a piece of metal   by studying its thermodynamic properties in 
different phases which can vary depending on whether   phonons or electrons  make the dominant contribution.

With this motivation, our objective  will be to develop the  \textit{macroscopic} computation of exact  quantum entropy 
for supersymmetric black holes in a broad class of phases of string theory, namely vacua with $\CN=2$ 
supersymmetry in four dimensions. In a theory with massless $n_{v} +1$ vector fields, a black hole is specified by a 
charge vector $(q_{I}, p^{I})$  with $I = 0, \ldots, n_{v}$. We would like to develop methods to systematically compute 
the quantum entropy for arbitrary finite values of the charges.   For this purpose, we use the  quantum entropy 
formalism developed by  Sen \cite{Sen:2008yk, Sen:2008vm} which generalizes the notion of Wald entropy 
\cite{Wald:1993nt, Iyer:1994ys, Jacobson:1994qe} in the case of extremal black holes using holography in the near 
horizon $AdS_{2}$ background.   In this formalism,   for a supersymmetric black hole of charge vector $(q, p)$, the  
macroscopic analog of the microscopic degeneracy $d(q, p)$ is given by the expectation value  $W(q, p)$ 
of a Wilson line  on the boundary of the near horizon $AdS_{2}$  with specific supersymmetric boundary conditions. 
Exact quantum entropy is thus given by a formal functional integral over all string fields on the $AdS_{2}$ 
background. We  review this formalism in \S{\ref{QuantumEntropy}}.

Evaluating the formal expression for $W(q, p)$ by doing the string field theory functional integral is of 
course highly nontrivial. To proceed further we  imagine first integrating out the infinite tower of massive string modes  
and massive Kaluza-Klein modes to obtain a \emph{local} Wilsonian effective action for the massless supergravity 
fields.  To compute the exact quantum entropy,  one has to then evaluate exactly this functional integral of a finite 
number of  massless fields with $AdS_{2}$ boundary conditions using the full Wilsonian effective action  keeping all 
higher derivative terms. This effective action can include in general not only perturbative corrections in $\alpha'$ but 
also  worldsheet instanton corrections.
We can  regard the ultraviolet finite string theory as providing a finite, supersymmetric, and consistent cutoff at 
the string scale. The functional integral with such a finite cut-off and a Wilsonian effective action containing all higher order terms is  thus in principle free of ultraviolet divergences. This functional integral will be our starting point.

We are still left with the task of evaluating a complicated functional integral.
The near horizon geometry preserves eight superconformal symmetries and moreover the  action, measure, 
operator insertion, boundary conditions of the functional integral are all supersymmetric. 
This allows us  to apply localization techniques \cite{Banerjee:2009af} which simplifies the evaluation of the functional integral enormously. Localization requires identification 
of a fermionic symmetry of the theory that squares to a compact bosonic symmetry.  Using this symmetry, one can 
then localize the functional integral onto the `localizing submanifold' of  bosonic field configurations invariant under the 
fermionic symmetry.  We review  the  superconformal symmetries of the near horizon geometry and  relevant aspects
of localization in  \S{\ref{Localization}. 

Since localization is employed at the level of the functional integral and not just at  the level of a classical action,  it is 
important to use an \textit{off-shell} formulation of supergravity. Off-shell formulations of supergravity are in general 
notoriously involved. At present a complete formulation of off-shell supergravity coupled to both vector and hyper 
multiplets is not known. 
To implement localization in a concrete manner,  we therefore first  consider in \S\ref{Solution} a simpler problem  of  computing this 
expectation value of the Wilson line  in a truncated model of   supergravity  coupled only to vector multiplets  with an 
action containing only  F-terms which are chiral integrals over superspace.  In particular we ignore possible D-terms and  hyper multiplets, which are discussed later in \S\ref{Connection}.  The action still contains an infinite number of higher derivative terms but all of F-type.
We denote the corresponding functional integral  for the expectation value of a Wilson line in this restricted 
theory on $AdS_{2}$ by $\hat W(q, p)$.   
Computation of $\hat W(q, p)$ is greatly simplified by the fact that,  for vector multiplets 
in $\CN=2$ supergravity,  there exists an elegant off-shell formulation developed  in 
\cite{deWit:1979ug, deWit:1984px, deWit:1980tn},  using the 
superconformal calculus. The spectrum consists of the Weyl multiplet that contains the graviton and the gravitini, 
$n_{v} + 1$ vector multiplets, and one compensating multiplet that eliminates unwanted degrees 
of freedom. We review this formalism in \S{\ref{off-shell}}.  

The main result of the paper concerns the localization  of the functional integral for $\hat W (q, p)$ which is  
derived in \S{\ref{Solution}}.   We now summarize the salient features  of this computation and  the  answers that we 
have been able to obtain. 
\begin{myenumerate}

\item  We choose a particular linear combination $Q$ of the superconformal superymmetries of the near horizon 
geometry  and deform the functional integral by adding to  the action a Q-exact term of the form 
$\lambda QV$ for $V = (Q\Psi, \Psi)$ where $\Psi$ refer to all fermionic fields of the theory and $\lambda$ is a 
continuous parameter that we introduce. We discuss the localizing action $S^{Q} :=QV$ in detail in 
\S{\eqref{Offshell}}.  As we will see, using off-shell supersymmetry variations is not only important for conceptual reasons but will turn out to be essential in this problem. In particular  some auxiliary fields which are set to zero on the on-shell theory will play a 
critical role and develop a nontrivial position dependence for the localizing solutions.

\item By the usual arguments of localization reviewed in \S{\ref{LocReview}}, 
the functional integral  
localizes onto the  critical points of the functional $S^{Q}$. We obtain  a family of nontrivial   instantons as 
exact  solutions to the equations of motion that follow from extremization of $S^{Q}$.  Since we use off-shell 
supersymmetry variations, these instanton solutions are completely \textit{universal }and independent of the form of the physical action.

\item For these solutions, the scalar fields $X^{I}$ in the vector multiplets are no longer fixed at the attractor values $X^{I}_{*}$ but 
have a nontrivial position dependence in the interior of the $AdS_{2}$ given by
\begin{equation}
X^{I} = X^{I}_{*} + \frac{ C^{I}}{\cosh (\eta)} \, , \quad \bar X^{I} = \bar X^{I}_{*} + \frac{ C^{I}}{\cosh (\eta)} 
\end{equation}
in coordinates where the metric on $AdS_{2}$ takes  form $ds^{2 } = \sinh^{2}(\eta) d\theta^{2} + d\eta^{2}$. 
 Auxiliary  fields in the vector multiplets are excited in such a way 
that the Q supersymmetry  is  preserved.  The family of solutions is  parametrized by  real parameters 
$\{C^{I}\}, I = 1, \ldots, n_{v} +1 $, which correspond to the values of the 
auxiliary fields in the vector multiplets at the center of $AdS_{2}$.  The infinite-dimensional functional integral 
thus localizes onto a finite number of ordinary bosonic integrals over the $\{C^{I}\}$.

\item  Many D-terms are expected to evaluate to zero on these solutions because of the nonrenormalization theorem of \cite{deWit:2010za}. We assume this to be true more generally and restrict our attention to F-type action\footnote{If this assumption is not true then the contribution of the D-terms can be systematically taken into account by simply evaluating these terms on  our localizing solution.}. 
Such actions are completely specified by specifying a single holomorphic 
prepotential ${F} (X^{I}, A)$ where $X^{I}$ are scalars in the vector multiplet and $A$ is the auxiliary field in the Weyl 
multiplet. For such actions, we evaluate the renormalized action $S_{ren}$  for the Q-invariant  localizing instanton configurations  exactly as a function of 
the $\{C^{I}\}$ following the prescription in \cite{Sen:2008yk, Sen:2008vm}.  This  action takes the form
\begin{eqnarray}
 \mathcal{S}_{ren}(\phi, q, p) =    - \pi q_I\phi^I  + \mathcal{F}(\phi, p ) \,
\end{eqnarray}
with $\mathcal{F}$ given by
\begin{equation} \label{freeenergy}
\mathcal{F}(\phi, p) = - 2\pi i \left[ F\Big(\frac{\phi^I+ip^I}{2} \Big) -
 \bar{F} \Big(\frac{\phi^I- ip^I}{2} \Big) \right] \, .
 \end{equation}
Note that $S_{ren}(\phi, q, p)$
equals precisely the 
\textit{classical} entropy function $\mathcal{E}( e, q,  p)$ \cite{Sen:2008yk, Sen:2008vm} but with electric fields replaced by the linear combination $
\phi^{I} := e_{*}^{I} + 2 C^{I}$ where $e_{*}^{I}$ are the attractor values of the electric field. 
\item 
We would like to emphasize that even though $S_{ren}$ and $\CE$   have the same functional form, their physical origin and meaning is very different. The entropy function is essentially a classical and on-shell object. Only its  extremum which determines the classical attractor values and its value at the extremum which determines the Wald entropy have physical meaning. By contrast, $S_{ren}$ is an intrinsically off-shell object valid for values of the fields far away from the classical attractor values.   Now, as long as  $S_{ren}$  has the same extremum and the value at the extremum as $\CE$ but differs from $\CE$,  it would reproduce the semi-classical results. Hence, a priori,  away from the extremum the two functions could have been  very different. It is thus something of a surprise that the nontrivial computation in \S\ref{Solution} for the off-shell instantons yields the same answer as the classical entropy function.

 \item The infinite dimensional functional integral thus reduces to the following finite dimensional integral
\begin{equation}\label{integral}
 \hat W (q, p) = \int_{\mathcal{M}_{Q}}   e^{ -\pi  \phi^{I} q_{I}}  \, e^{\mathcal{F}(\phi, p)}
 \,  \, |Z_{inst}|^{2} \, Z_{det}\,  [dC]_{\mu}
\end{equation}

The measure of integration over $[dC]_{\mu}$ is computable from  the original measure $\mu$ of the functional integral of massless fields of string theory by standard collective coordinate methods. The factor $Z_{det}$ are the one-loop determinants of the quadratic fluctuation operator around the localizing instanton solution. Such one-loop determinant factors in closely related problems have been computed in \cite{Pestun:2007rz, Banerjee:2010qc}. This computations are straightforward in principle but technically involved and we defer their detailed discussion to a forthcoming publication \cite{Dabholkar:2010t}. 

\item 
We have included a term  $|Z_{inst}|^{2}$ to indicate the contribution of brane-instantons which in general will be present 
in the full string theory computation. Supersymmetric configurations in $AdS_{2} \times S^{2}$ typically  correspond to instantons localized at the north pole of $S^{2}$ with counting function $Z_{inst}$ and anti instantons localized at the south pole with counting function $\bar Z_{inst}$ \cite{Beasley:2006us, Pestun:2007rz}.  Note that  nonperturbative instantons in one duality frame may be incorporated as worldsheet instantons in another duality frame, and hence the separation between the classical piece and instanton piece is frame-dependent.   We keep it general at this stage to underscore the fact that the physics of black hole horizons is also frame-dependent. This is a consequence of the fact that for the same asymptotic states,  some degrees of freedom that are  horizon degrees of freedom in one frame may be  external to the horizon in another frame \cite{Banerjee:2009uk, Sen:2009bm}. \end{myenumerate}

Independent of our original motivation of computing quantum entropy of black holes, results in  section 
\S{\ref{Solution}} could be viewed simply as results about the localization of this functional integral of supergravity 
coupled to vector multiplets keeping only F-terms. Equation \eqref{integral} represents  a remarkable reduction of a 
complicated functional integral of gravity onto an ordinary integral with an integrand that depends on the prepotential in a particularly simple way.

In \S{\ref{Connection}}, we proceed to the evaluation of   $W(q, p)$ using these results for $\hat 
W(q, p)$. For this purpose it is necessary to make  the assumption that the hyper multiplets and D-terms 
do not contribute. We examine this in some detail  in \S\ref{Eval} in  view of the nonrenormalization theorem of \cite{deWit:2010za}.  We then discuss in \S\ref{nonpert}
the  subleading gravitational saddle points which will in general give nonperturbative contributions in addition to those from brane-instantons.
We discuss how these saddle point can be taken into account in a general setup. 
We conclude \S\ref{Relation} with comments about the connection with the topological string and  the Donaldson-Thomas invariants. 
Before summarizing the general structure of the localization integral, we would like to  clarify two common misconceptions.
\begin{myitemize}
\item \textsl{Choice of the ensemble:}
The large area limit is the thermodynamic limit. The leading answer for the entropy in this limit is the Bekenstein-
Hawking entropy which does not depend on the choice of the ensemble. However, the subleading corrections, which 
are the finite size effects, depend sensitively on the ensemble. It is thus necessary to determine which ensemble is 
natural on the macroscopic side. The natural ensemble from the  perspective of the $AdS_{2}$ boundary conditions   
\cite{Sen:2008yk, Sen:2008vm}  is the microcanonical 
one\footnote{In some cases it is possible that both boundary conditions are
physically allowed as in the Breitenlohner-Freedman window in higher
dimensions. The two choices of the  boundary conditions lead to physically distinct
theories whose partition functions could be regarded to be related to
each other by a Legendre transform \cite{Klebanov:1999tb}. Thus it may be
possible to use the mixed ensemble in special cases, especially to compare with the OSV conjecture \cite{Ooguri:2004zv} where the mixed ensemble arises naturally. However, in generic examples  the sum over charges  
may not be convergent and it is not clear if  the mixed ensemble can be  defined  beyond an asymptotic expansion.  For a detailed discussion of the comparison of the two ensembles see \cite{Sen:2008vm}.  In this paper we use the  microcanonical boundary conditions which are most
general and natural in the $AdS_{2}$ set-up.}. 
This follows from the fact that in $AdS_{2}$, the Coulomb potential grows linearly at infinity. 
As a result,  in $AdS_{2}$, the fluctuation of charge corresponds to a  
nonnormalizable mode unlike say in $AdS_{5}$, and must be held fixed at the boundary. 
\item \textsl{Index vs degeneracy:}
The entropy of a black hole is a thermodynamic quantity and accoding to Boltzmann relation it should be compared 
with the absolute number of states.  The supersymmetric techniques that we will be using  would seem more 
appropriate for the computation of an index and not the entropy or the  absolute number. However, for a black hole 
with an $AdS_{2}$ near horizon geometry that preserves at least four supersymmetries, the index equals the 
absolute number. This follows from the following reasoning \cite{Sen:2009vz}. Demanding closure of the algebra containing the 
$SU(1, 1)$ isometries of the $AdS_{2}$ together with the four supersymmetries implies that the symmetry is a 
supergroup $SU(1, 1|2)$ which contains an $SU(2)$ symmetry. This implies that the black hole horizon has rotational symmetry\footnote{The well-known fact that there are no spinning supersymmetric black holes in four dimensions is a 
consequence of this fact.}.  
In a thermodynamic ensemble,  this could either mean that  its spin $J$ is zero or the chemical potential $\mu$ 
conjugate to the spin is zero. However, the boundary conditions for the choice of the ensemble discussed above 
require that we hold $J$ fixed which appears as a charge in $AdS_{2}$ and not the chemical potential. This implies that for the microstates associated 
with the horizon of a single black hole
\begin{equation}
\textrm{Tr} (-1)^{F} := Tr (e^{2\pi i J}) = Tr (1) \, ,
\end{equation}
and hence the index equals the absolute number. Since our computation will be near the horizon of a single black 
hole, the argument above shows that it is justified to use supersymmetric localization methods even for the 
computation of the absolute number. If one wishes to compare this with a microscopic computation which is often 
done for an index at asymptotic infinity it is necessary to define an analogous index on the macroscopic side using our 
results here as an input. For a more detailed discussion see \cite{Dabholkar:2010rm}.
\end{myitemize}

Putting these various ingredients together, the final answer for the quantum entropy  is  of the general form
\begin{equation}\label{Wsum}
W(q, p) = \sum_{s} W_{s}(q, p) \ ,
\end{equation}
where $s$ is a non-negative integer labeling  various  orbifolds with $s=0$ being the unorbifolded theory. The term $W_{0}$ denotes  the functional integral on unorbifolded $AdS_{2}$ which gives  the leading contribution and will be discussed in \S\ref{Eval}. In the semilclassical limit, $W_{0}$ will scale as $\exp{(S_{Wald})}$ whereas $W_{s}$ for nonzero $s$ will scale as  $\exp{(S_{Wald}/c)}$ where $c$ is the order of the orbifold. These terms are thus exponentially subleading compared to $W_{0}$ and will be discussed in \S\ref{nonpert}. 
We would like  to emphasize two features of this answer.

 \begin{myenumerate}
\item 
 If D-terms can be ignored for reasons of supersymmetry, then we expect  $W_{0}$ will equal $\hat W$ given by \eqref{integral}. In the Type-IIA frame, the world sheet instanton corrections are incorporated in the prepotential computed by the topological string. Moreover, it is likely that there are no spacetime instanton  contributions from wrapped D-brane or NS5-branes because their action would have to depend on the dilaton which is in the hyper-multiplet. In  other words, such instantons wrapping an internal cycle of the Calabi-Yau but localized at the poles of $S^{2}$ may not contribute to the supersymmetric functional integral. We have not analyzed this in detail but if true, it would imply that 
 \begin{equation}\label{integral1}
 W_{0} (q, p) = \int_{\mathcal{M}_{Q}} \, e^{ -\pi  \phi^{I} q_{I}}  \, | Z_{top}(\phi, p ) |^{2}  \, 
 \, Z_{det}  \, [dC]_{\mu} \ . 
 \end{equation}
The  term $|Z_{top}|^{2}$ is precisely of the form envisaged by Ooguri-Strominger-Vafa \cite{Ooguri:2004zv}! It should be emphasized though that  in this derivation, the topological string partition function  in \eqref{integral1} is evaluated for  the values of the scalar fields at the \textit{center} of $AdS_{2}$ and not at the boundary of $AdS_{2}$. Since the scalar fields are moving away from the attractor values, they are no longer at the minimum of the entropy function and hence are thus `climbing up' the entropy function away from the critical point.
\item  As mentioned above, for a consistent treatment of the $AdS_{2}$ path integral it is necessary to work in the microcanonical ensemble because in $AdS_{2}$ Coulomb potential is a a nonnormalizable mode. Various fields approach their attractor values at the boundary in this fixed charged sector. In particular,  the electric fields at the boundary do not fluctuate and  remain fixed at their attractor values $e_{*}$.  For our localizing instanton solutions the fields move \textit{off-shell} away from the attractor values inside $AdS_{2}$ and it is the value of these fields at the center of $AdS_{2}$ that is allowed to fluctuate and can be integrated over  as in \eqref{integral1}.  This derivation thus requires off-shell supergravity in an essential way. 
\item It has been pointed out that agreement with statistical entropy requires modifying the OSV conjecture by including additional measure factors  and nonperturbative corrections \cite{Dabholkar:2005by, Dabholkar:2005dt, Shih:2005he,LopesCardoso:2006bg, Denef:2007vg, Cardoso:2008ej}. Our localizing solution provides a well-defined starting point to define these corrections systematically on the macroscopic side  as we discuss below.   See \S\ref{Relation} for a more detailed discussion of the relation to the topological string and  comparison with  earlier work.
\item Note that the range of integration is determined by the localization analysis and goes from $-\infty$ to $+\infty$ for all $C$. Since the measure factor and $Z_{det}$ follow algorithmically from the original measure of the string theory spacetime fields, the resulting answer is guaranteed to respect all symmetries.
In particular, even though the renormalized action is Wilsonian and given in terms of holomorphic prepotential,  one expects from the $AdS_{2}/CFT_{1}$ correspondence that holomorphic anomalies \cite{Bershadsky:1993ta, Bershadsky:1993cx, Antoniadis:1993ze} are taken into account systematically by the integral \eqref{integral1} so that the resulting degeneracies respect all duality symmetries. 
\item The higher orbifold contributions in \eqref{Wsum} appear to play an important role if the macroscopic analysis is to reproduce the known microscopic answers in the $\CN=4$ examples \cite{Dabholkar:2010t}. In particular, their contribution would be essential to ensure that the resulting sum is an integer.  {}From this point of view as well, it is more natural to work in the microcanonical ensemble than invert \eqref{integral1} and work in a mixed ensemble.
\end{myenumerate}

To summarize, in this paper we have solved one major piece  of the puzzle in the evaluation of the black hole functional integral. We have determined  which off-shell field configurations to integrate over consistent with the $AdS_{2}$ boundary conditions.  We have obtained  explicit analytic expressions for the  localizing instanton solutions and the renormalized action evaluated on these instantons.  These solutions thus give a well-defined starting point to evaluate various corrections to the semi-classical results. Further work is necessary to determine one-loop contributions and the measure \cite{Dabholkar:2010t}. This computation is technically involved but we would like to emphasize that given the localizing action and the localizing instantons this step is essentially algorithmic. Computation of brane-instantons in general case will be more complicated since it will depend on the details of the compactification but can simplify in specific examples and in specific duality frames.
 
Finally, we note that in models where exact microscopic degeneracies  $d(q, p)$ are known from independent computations,  our 
macroscopic computation can be tested against the microscopic answer.  Such a comparison for half-BPS small black 
holes and quarter-BPS big black holes in $\CN=4$ compactifications  will be reported in a forthcoming publication \cite{Dabholkar:2010t}. In 
these examples,   all the ingredients of the macroscopic computation discussed in this paper play an important role. 
With $\CN=4$ supersymmetry,  all three terms in the integrand $\exp (\CF )$, $Z_{det}$ and $Z_{inst}$ can be  
computed almost completely. Moerover, the resulting integral representation can be immediately be put a form that 
that agrees in remarkable detail with the microscopic degeneracies including all perturbative and nonperturbative 
corrections. These results can thus be seen as a partial post-facto justification for some of the assumptions made in 
section \S{\ref{Connection}}.

\section{Quantum entropy and $AdS_{2}/CFT_{1}$ correspondence \label{QuantumEntropy}}

The quantum entropy formalism \cite{Sen:2008yk, Sen:2008vm}  generalizes the Wald entropy formula 
to include 
quantum corrections to black hole entropy in a consistent quantum theory of gravity such as string theory. 
It is formulated  in general for any extremal black holes whose near horizon geometry has an  $AdS_{2}$  factor. We would now like to apply this formalism to four-dimensional  supersymmetric black hole whose  near-horizon geometry
is of the form $AdS_{2} \times S^{2} \times K$ where $K$ is a  three complex dimensional 
Calabi-Yau manifold of string compactification.  

Since the essential physics of the quantum entropy concerns the $AdS_{2}$ factor, for the rest of this section  we will dimensionally reduce all the way to two dimensions onto $AdS_{2}$.  
One can regard the full theory as a two-dimensional theory of gravity interacting with an infinite set of fields keeping all massive modes.
The massless sector consists of the 2D metric, a set of gauge fields $A^{i}$ with field strengths $F^{i}$, 
and matter fields $\phi^{a}$ which include the moduli of $K$, as well as the fluxes through 
the various cycles in the `internal' geometry $S^{2}\times K$. 
The electric charge of the four-dimensional black hole is represented by the gauge fields, and the magnetic charges which correspond to fluxes through the $S^{2}$ are represented as  fixed parameters  of the theory living on the $AdS_{2}$ geometry.

\subsection{Near horizon classical geometry}

The most general near horizon configuration for the massless fields consistent with the 
$SL(2,\IR)$ isometry of $AdS_{2}$ is:
\be\label{nearhor}
 ds^{2} = v \left[ -(r^{2}-1) d t^{2} + \frac{d r^{2}}{r^{2} -1}   \right] \ ,\qquad 
F^{i} = e^{i} d r \wedge d t\ , \qquad \phi^{a} (t ,r)= \phi^{a}_0 \ ,
\ee
where $v, e^{i}$ and $\phi^{a}_0$ are constants. 
This is the metric of an $AdS_{2}$ black hole \cite{Jackiw:1982hg, Jackiw:1984je, 
Teitelboim:1983fg, Spradlin:1999bn} 
with horizon at $r = 1$. It is locally isometric to $AdS_2$ and the region $r>1$
covers a triangular wedge extending halfway from the boundary 
into global $AdS_2$ \cite{Spradlin:1999bn}. 
An analytic continuation $ t = - i \theta $ leads to the Euclidean metric
\be\label{nearhoreucl}
ds^{2} = v \left[ (r^{2}-1) d\theta^{2} + \frac{d r^{2}}{r^{2} -1}  \right]\ , \qquad 
F^i= -i\, e^{i} d r \wedge d \theta, \qquad \phi^{a} (\theta,r)= \phi^{a}_0\ .
\ee
This metric is non-singular at the erstwhile horizon $r =1$ 
provided  the Euclidean time
coordinate $\theta$ is periodic modulo $2 \pi$. In the gauge $A^i_{r} =0$, the gauge 
fields  are given by  
\be\label{ggefld} 
A^{i} = -i \, e^{i} (r -1) d \theta \, ,
\ee
where the constant term ensures that the Wilson line $\oint_{S^1} A^i$ around the 
thermal circle vanishes at the horizon $r=1$. This is needed for regularity since
the thermal circle contracts to zero size. 

It is worth emphasizing that it is important to use the form of the metric in \eqref{nearhor} with two separate horizons at $r = \pm 1$
which corresponds to the Jackiw-Teitelboim black hole \cite{Jackiw:1982hg, Jackiw:1984je, 
Teitelboim:1983fg}. Physically this corresponds to staying close to the black hole horizon as one takes the near horizon limit in which the $AdS_{2}$ throat becomes infinitely long.  Upon Euclidean continuation this covers the entire upper half plane or the Poincar\'e disk which has the topology of a disk and hence Euler character one. In the Gibbons-Hawking formalism, the entropy of the black hole is proportional to the Euler character of the near horizon in the $(r, t)$ plane and hence one obtains finite entropy. If we use instead the metric
\be\label{nearhor2}
 ds^{2} = v \left[ \rho^{2} d \theta^{2} + \frac{d 
 \rho^{2}}{\rho^{2}}   \right] \ ,\ee
with periodic $ \theta$ then this  covers only a strip in the upper half plane  with two edges 
identified. The geometry then has a topology of a cylinder, or a punctured disk, which 
has Euler character zero, and hence vanishing entropy \cite{Hawking:1994ii}. For applications in string 
theory it has been clear that an extremal black hole should really be thought of as a limit of a non-extremal black hole in which case it has zero temperature but nonzero entropy. This corresponds to choosing the metric as in \eqref{nearhor}. 

\subsection{Functional integral for the quantum entropy}

The quantum entropy  is defined by  a functional integral over all field
configurations which asymptote to the
$AdS_{2}$ Euclidean black hole \eqref{nearhoreucl} with the fall-off conditions \cite{Castro:2008ms}
\bea\label{asympcond}
d s^2_{0} &=& v \left[ 
\left(r^2+\cO(1)\right) d\theta^2+ \frac{dr^2}{r^2+\cO(1)}  \right]\  ,\cr
\phi^a & =&u^a + \cO(1/r)\ ,\qquad
A^i = -i  \, e^{i} (r -\cO(1) ) d\theta\ ,
\eea
which are invariant under an action of the Virasoro algebra.
In particular, in contrast to higher dimensional instances of the AdS/CFT correspondence,
the mode of the gauge field corresponding 
to the electric  field   grows linearly (or is `non-normalizable') and must be kept fixed,
while the mode corresponding to the electric potential  is  constant (or is `normalizable'), and  
allowed to fluctuate. Since the asymptotic value of the electric field is determined by the charge of the black hole by Gauss law
\be
\label{qerel}
q_{i} = \frac{\p (v \CL)}{\p e^{i}} \ ,
\ee
this is equivalent to holding the charge fixed. 
The asymptotic values of the parameters 
of the metric and the scalars are determined purely in terms of the charges by the 
{\it attractor mechanism}. 
The constants $v,e^{i}, u^{a}$ which set the boundary conditions of the path 
integral must therefore be set to their  attractor  values $v_{*},e^{i}_{*}, u^{a}_{*}$ respectively. 
The quantum entropy is thus purely a function of the electric charges $q_i$.
This was defined by \cite{Sen:2008vm} as the functional integral with an insertion of the Wilson line:
\be\label{qef}
W (q, p) = \left\langle \exp \big[-i \, q_i \int_{0}^{2 \pi}  A^i_{\theta} \, d \theta \big]  \right\rangle_{\rm{AdS}_2}^{\rm finite}\ .
\ee
Note that in the classical limit, this constant mode of the gauge field gets  determined in terms of  the attractor electric field $e^I_{*}$ by the smoothness condition on the classical gauge field but in the quantum theory it is free to fluctuate.

%

We will now explain the meaning of the superscript \emph{finite} in the above functional integral. 
The action entering into this functional integral is of the form
\be\label{pathintwt}
S_{\rm bulk} + S_{\rm bdry}   \ ,
\ee  
where the actions\footnote{The signs in the first equation for the bulk action is chosen in accord with the 
Euclidean continuation from the Minkowski theory \cite{Sen:2008yk}, the sign in the second equation 
is a convention since it is a one dimensional Euclidean problem, which we have fixed.} 
\be\label{Avalue}
S_{\rm bulk} =  \int \, \CL_{\rm bulk} \, \sqrt{g} \,  dr \, d \theta \ , 
\qquad S_{\rm bdry} =   \int  \CL_{\rm bdry}  \, \sqrt{g_{\rm ind}} \,  d \theta \ 
\ee
are expressed in terms of local Lagrangian densities, the measure in the boundary term 
coming from the induced metric on the boundary. 
The integral for the bulk action over $r$ suffers from an obvious infrared divergence due to the infinite volume of the $AdS_{2}$. 
The superscript {\it finite} in \eqref{qef} refers to the following prescription for regulating 
and renormalizing this divergence. 

First, one enforces a cutoff at a large $r = r_{0}$. This cutoff which respects the angular 
symmetry seems to be special, but the conclusions below have been shown to be 
independent of the details of the cutoff \cite{Sen:2009vz}. 
The bulk Lagrangian density $\CL_{bulk}$ is the full local classical Lagrangian of the theory 
including all massive fields. 
Since $\CL_{bulk}$
is a local functional of the fields,  the bulk effective action evaluated on a certain field configuration 
has the form 
\be\label{r0Taylor}
S_{\rm bulk} = C_{0} r_0 + C_{1} + \cO(r_0^{-1}) \ , 
\ee
with $C_{0}, C_{1}$ independent of $r_{0}$.
 
The boundary action is the boundary Lagrangian $\CL_{\rm bdry}$ multiplied by the the proper length 
$L \sim 2 \pi \sqrt{v} r_{0}$ of the boundary which goes to infinity as $r_{0} \to \infty$. 
$\CL_{\rm bdry}$ is a local gauge invariant functional of the fields of the theory. 
Using the asymptotic form of the fields \eqref{asympcond}, one obtains that  
the boundary action has a form like \eqref{r0Taylor} with coefficients that 
only depend on the classical values of the various fields in the problem which are held fixed.

One now chooses the boundary counterterms such that the piece linear in $r_{0}$ in the integrand 
vanishes, as is standard in the procedure of holographic renormalization. 
In particular, one can subtract the constant piece $C_{0} r_{0}$ from the action simply by using 
an appropriate boundary cosmological constant as in \cite{Sen:2008vm}. 
This ensures that the boundary Hamiltonian of the $CFT_{1}$ dual to $AdS_{2}$ is zero.
From the above argument, this boundary cosmological constant is a function of the classical 
values of the fields. There can be of course other finite parts of the boundary action which 
depend on the fluctuating parts of the fields, this will be part of the full definition of the quantum 
entropy function. We shall comment on them below. 

 It is convenient to incorporate also the Wilson line into the renormalized action and include counterterms to cancel  the divergences in the Wilson line.
One can then take the limit $r_{0} \to \infty$, and define the finite part of the path integral 
unambiguously as $e^{-S_{\rm ren}}$. We  refer to this finite piece $S_{\rm ren}$ as the renormalized action, 
which in general is a functional of all the fields.  We thus have the definition
\be\label{Sren}
\CS_{\rm ren} :=    \CS_{\rm bulk}  + \CS_{\rm bdry} - i \, {q_i }  \int_{0}^{2\pi} A^i_{\theta}  \, d\theta \ .
\ee

In the classical limit, the functional integral \eqref{qef} is dominated by the saddle point 
where all fields take their classical values \eqref{nearhoreucl}. In this case, the path integral reduces to 
\be\label{classlim1}
\left\langle \exp\big[-i \, q_i \int_{0}^{2\pi} A^i d \theta \big]  \right\rangle = \exp{ \left( S_{\rm bulk} + S_{\rm bdry} - i q_{i} \int_{0}^{2\pi} A^i_{\theta} d \theta  \right)}  \ ,
\ee  
where $S_{\rm bulk}$ and $S_{\rm bdry}$ are as above. In this case, one can simply evaluate the bulk Lagrangian 
at the constant classical values \eqref{nearhor} to get  
\bea\label{Avalue2}
S_{\rm bulk} & = &  \int_{0}^{2\pi} d \theta \int_{1}^{r_{0}} dr \, v \, \CL = 2 \pi \, (r_{0} - 1) \, v \,  \CL 
 \ , \\ 
S_{\rm bdry}  & = & - 2\pi  r_{0} \, ( v \mathcal{L} - q_{i }e^{i})+ \CO(1/r_{0}) \ , \qquad  - i q_{i}  \int_{0}^{2\pi} A^i_{\theta} d \theta  = -  2  \pi \, q_{i} e^{i} \, (r_{0}-1) \ . 
\eea
After the above regulation and renormalization procedure, one has 
\be\label{classlim0}
W(q, p) \sim  \exp[{2 \pi ( q_{i} e^{i} - v \CL)}]  \equiv \exp\left[{S_{Wald}(q, p)}\right] \ ,
\ee  
where it is understood that the middle term is evaluated at the attractor values of the fields. 
Since the attractor values of various fields and in particular the electric fields are  determined by extermization of the classical action,  one can define the entropy function 
\begin{equation}
\CE(e, p, q) := 2 \pi ( q_{i} e^{i} - v \CL (e, p)) \, .
\end{equation}
Here we have fixed the scalars to their attractor values but kept the electric fields as variables. By virtue of its construction, the classical attractor values $e_{*}(q, p)$ of the electric fields can be found at the extremum 
of $\CE$  which are determined entirely in terms of the charges. As shown in \cite{Sen:2008yk}, 
the value of the entropy function at the extremum  equals the Bekenstein-Hawking-Wald entropy of the extremal black hole.

In the case of supersymmetric black holes, we should use a supersymmetric version 
of this Wilson line which requires the addition of another boundary term to the Wilson line. 
We find that the bulk Lagrangian is also supersymmetric only up to a boundary term which 
has to be cancelled by adding a boundary counterterm. Both these additional boundary terms are field 
dependent. However, it turns out that the sum of the two additional boundary 
terms is equal to a constant diverging linearly with $r_{0}$, and moreover, 
it is precisely the constant required to cancel the infrared divergence of the original bulk action plus Wilson line.
As a result we can express the total functional integral in a manifestly supersymmetric manner, at the same 
time using the naive operational definition above. 
We will discuss these issues and the supersymmetry of the functional integral in more detail in 
\S\ref{RenAction} and \S\ref{Supersymmetry}.

In the full quantum theory, there will be corrections to the classical answer which are of two kinds. 
The first type of correction will arise from evaluating the functional integral around the $AdS_{2}$ geometry. 
One can try to evaluate it in a saddle point approximation, but this can at best give an asymptotic 
expansion and one can never access large fluctuations in field space. We shall overcome this using 
the technique of localization in the context of supersymmetric theories. This will allow us to  
evaluate the functional integral in the  $AdS_{2}$ background exactly. We discuss this in the next sections. 

The second type of correction comes from subleading orbifold saddle points that play an important role. 
Keeping these subleading saddle points which are much smaller than the power law corrections 
to the leading saddle point cannot be justified in an asymptotic expansion, but the exact 
evaluation of the functional integral allows us here  to consistently deal with subleading saddles. 
The full functional integral localizes onto a discrete series, and for each term in the series we 
obtain an exact finite dimensional integral which accesses large fluctuations in field space.
We  discuss the general form of the localizing integral in \S\ref{nonpert}.

\section{Superconformal symmetries and localization \label{Localization}}

We start with a  brief review in \S\ref{LocReview} of  the localization techniques 
\cite{Duistermaat:1982xu,Witten:1988ze,Witten:1991mk,Witten:1991zz,Schwarz:1995dg,Zaboronsky:1996qn}
 to evaluate supersymmetric functional integrals. In \S\ref{Supersymmetries} we review the superconformal  symmetries  of the attractor geometry and how localization can be applied in the present context.

\subsection{A review of localization of supersymmetric functional integrals \label{LocReview}}

Consider a supermanifold $\mathcal{M}$ with an integration measure $d\mu$. Let $Q$ be an odd (fermionic) vector field on this manifold that satisfies the following two requirements:
\begin{myitemize}
\item $Q^{2} =H$ for some compact bosonic vector field $H$,
\item The measure is invariant under $Q$, in other words $div_{\mu} Q = 0$.
\end{myitemize}
The divergence of the fermionic vector field is the natural generalization of ordinary divergence, which satisfies in particular\footnote{For a bosonic vector field $V$ and for  a measure determined by a metric $g$,  this corresponds to the identity   $\int dx \sqrt{g} V^{m} \partial_{m}f = - \int dx \partial_{m}(\sqrt{g} V^{m}) f  = - \int dx \sqrt{g} (\nabla_{m} V^{m})  f$  when the boundary contributions vanish.}
 \begin{equation}
\int_{\mathcal{M}} d\mu \, Q (f) = - \int_{\mathcal{M}} d\mu (div_{\mu}Q)  \, f \, ,
\end{equation}
for any function $f$. Hence, the second property implies $\int_{\mathcal{M}} d\mu \, Q (f) = 0$ for any $f$. We would like to evaluate an integral of some $Q$-invariant function $h$ and a Q-invariant action $S$
\begin{equation}
I := \int_{\mathcal{M}} d\mu  \, h \, e^{\mathcal{- S}} .
\end{equation}
To evaluate this integral using localization, one first deforms the  integral to
\begin{equation}\label{exact deform}
I (\lambda)  = \int_{\mathcal{M}} d\mu  \, h \, e^{-\mathcal{S}  - \lambda QV} \ , 
\end{equation}
where $V$ is a fermionic, H-invariant function which means  $Q^{2} V = 0$ and  $Q V$ is Q-exact. One has 
\begin{equation}
\frac{d}{d\lambda}\int_{\mathcal{M}} d \mu   \, h  \, e^{- \mathcal{S} - \lambda QV} = \int_{\mathcal{M}} d \mu   \, h  \, QV \, e^{- \mathcal{S} - \lambda QV} = \int_{\mathcal{M}} d \mu   \, Q( h  \, e^{- \mathcal{S} - \lambda QV}) = 0 \ , 
\end{equation}
and hence $I(\lambda)$ is independent of $\lambda$. 
This implies that one can perform the integral $I(\lambda)$ for any value of $\lambda$ and in particular for $\lambda \rightarrow \infty$.  
In this limit, the functional integral  localizes onto the  critical points of the functional $S^{Q} := QV$ 
which we refer to as the localizing solutions. The localizing solutions in general have both bosonic and fermionic collective coordinates. 

One can choose 
\begin{equation}\label{locV}
V = (Q\Psi, \Psi)
\end{equation}
 where $\Psi$ are the fermionic coordinates with some positive definite inner product  defined on the fermions.
In this case, the bosonic part of  $S^{Q}$ can be written as a perfect square $(Q\Psi, Q\Psi)$, and hence critical points of $S^{Q}$ are the same as the critical points of $Q$. Let us denote this set of critical points of $Q$ by $\mathcal{M}_{Q}$. 
The reasoning above shows that the integral over the supermanifold $\mathcal{M}$ localizes to an integral over the submanifold $\mathcal{M}_{Q}$. 
In the large $\lambda$ limit, the integration for directions transverse can be performed exactly in the saddle point evaluation. One is then left with an integral over the submanifold $\mathcal{M}_{Q}$
\begin{equation}
I = \int_{\mathcal{M}_{Q}} d\mu_{Q} \, h \, e^{-\mathcal{S}} \, ,
\end{equation}
with a measure $d\mu_{Q}$ induced on the submanifold by the original measure. 

In our case in \S\ref{Solution}, $\mathcal{M}$ is the field space of off-shell supergravity, $\mathcal{S}$ is the off-shell supergravity action with appropriate boundary terms, $h$ is the supersymmetric Wilson line, $Q$ is a specific supercharge described in \S{\ref{Killing}} and \S{\ref{Solution}}, and $\Psi$ are all fermionic fields of the theory. 
We will find that the submanifold $\mathcal{M}_{Q}$ of localizing solutions is  a family of nontrivial   instantons as exact  solutions to the equations of motion that follow from extremization of $S^{Q}$ labeled by $n_{v} +1$ real parameters $\{ C^{I} \, ; \, I= 0, \ldots, n_{v}\}$.

\subsection{Superconformal symmetries of the near horizon geometry \label{Supersymmetries}}

The near-horizon geometry  of a supersymmetric black hole in four dimensions 
is $AdS_2\times S^2$. After Euclidean continuation,  the metric is
\begin{eqnarray}\label{ads2s2}
 ds^2=  v \left[(r^2-1)d\theta^2+\frac{dr^2}{r^2-1}\right] + v  \, \left[ d\psi^2 + \sin^2 (\psi) d \phi^2 \right]
  \,  . 
\end{eqnarray}
We have  taken the radius $v$  of the $AdS_{2}$ factor to be the same as the radius of the $S^{2}$ factor which is a consequence of supersymmetry. There are several other coordinates that are useful. Substituting $r = \cosh (\eta)$, the metric takes the form
\begin{equation}\label{metric2}
    ds^2 = v \left[ d\eta^2 + \sinh^2 (\eta) d\theta^2 \right] +  v\left[ d\psi^2 + \sin^2(\psi) d \phi^2 \right] \, . 
\end{equation}
One can also choose the stereographic coordinates 
\begin{equation}\label{coordinate-trans}
    w = \tanh (\frac{\eta}{2}) e^{i\theta} := \rho e^{i\theta}, \quad z = \tan (\frac{\psi}{2}) e^{i\phi} \ , 
\end{equation}
in which the metric takes the form
\begin{equation}\label{metric}
    ds^2 =v \frac{4 dw d\bar w}{(1- w \bar w)^2} + v  \frac{4 dz d\bar z}{(1 + z \bar z)^2} \, .
\end{equation}
Note that the interval for the coordinates are $ 1 \leq r  < \infty$ and $ 0 \leq \eta < \infty$, and $0 \leq \rho < 1$. In the $\w$ coordinates, 
Euclidean $AdS_{2}$ can be readily recognized as the Poincar\'e disk with
$\rho$   as the radial coordinate of the disk and a boundary at $\rho =1$. 

The Weyl tensor 
for the metric \eqref{metric2} is zero and hence this metric is conformally flat. For later use it will useful to know this conformal transformation. To map we first map the Poincar\'e disk to the upper half plane by  the transformation
\begin{eqnarray}
  u = x  + i y ,\, \qquad u = i \frac{1 - i w}{1 + iw} \, .
 \end{eqnarray}
The metric \eqref{ads2s2} in the new coordinates becomes
\begin{equation}\label{AdS_2 conformal flat}
 ds^2=\frac{dx^2+dy^2+y^2d\Omega_2^2}{y^2} \ , 
\end{equation}
with $-\infty < x < +\infty$ and $0  \leq y < \infty$. 
{}From the above equation,  we  see that $AdS_2\times S^2$ is conformally flat. We also know that $\mathbb{R}^4$ is conformal to $S^4$ so it would be useful to compute the conformal factor relating $AdS_2\times S^2$ to $S^4$. In the $(\eta,\theta)$ coordinates we have the following conformal rescaling
\begin{equation}\label{conformal-trans}
 ds^2(AdS_2\times S^2)=\cosh^{2}(\eta) ds^2(S^4) \ .
\end{equation}
Note that the conformal factor diverges at the boundary.
Under a Weyl transformation 
\be
 g_{\mu\nu} \rightarrow e^{2\Omega}g_{\mu\nu} \ , 
\ee
a field with Weyl weight $a$ 
transforms as 
\begin{equation}\label{scaling}
\Phi\rightarrow e^{-a\Omega}\Phi \ . 
\end{equation}
Hence, such a field in the conformal frame with $AdS_2\times S^2$ metric will be mapped  to the field 
in the conformal frame with $S^{4}$ metric by
\begin{equation}\label{Weylscaling}
 \Phi_{AdS_2\times S^2}=\frac{\Phi_{S^4}}{\cosh(\eta)^{a}}.
\end{equation}
This transformation will be useful later in \S{\ref{Solution}}.

The superconformal symmetry of the near horizon geometry is the semidirect product $SU(1, 1|2) \rtimes SU(2)'$.  The invariant subgroup
$SU(1, 1|2)$ will be of our main interest which contains the bosonic subgroup 
$SU(1, 1) \times SU(2)$. The first factor can be identified with the conformal 
symmetry of $AdS_2$ and is generated by $\{ L, L_\pm\}$. The second factor can
be identified with the rotational symmetry  of $S^2$ and is generated by $\{ J,
J_\pm\}$. The factor $SU(2)'$ originates from the R-symmetry of $\CN=2$ supergravity in four dimensions.   The odd elements of the superalgebra are the superconformal symmetries
$G^{ia}_r$. The commutations relations are
\begin{eqnarray}\label{scalgebra}
 \left[L, L_{\pm }\right] &=& \pm L_{\pm } \ ,  \qquad\qquad  \left[L_{+} ,
L_{-}\right] = -2 L  \ ,   \\
 \left[J, J^{\pm}\right] &=& \pm J^{\pm} \ ,   \qquad\qquad \left[J^+, J^-\right] = 2
J  \ ,   \\
 \left[L, G^{ia}_{\pm}\right] &=& \pm \half G^{ia}_{\pm} \ ,   \quad\qquad 
\left[L_\pm, G^{ia}_{\mp} \right] = -i  G^{ia}_{\pm} \ ,   \\
 \left[J, G^{i\pm}_r\right] &=& \pm \half G^{i\pm}_{r} \ ,   \quad\qquad \left[J^\pm,
G^{i\mp}_{r} \right] =   G^{i\pm}_{r}  \ ,  \\
 \{ G_+^{i\pm}, G_{-}^{j\pm}\} &=& \pm 4  \e^{ij} J^\pm \ ,   \qquad \{ G_\pm^{i+},
G_{\pm}^{j-}\} = \mp 4 i \e^{ij} L_\pm \ ,   \\
 &&\{ G_\pm^{i+}, G_{\mp}^{j-}\} = 4 \e^{ij} (L \mp J) \ ; \quad 
 \epsilon^{+-} = - \epsilon{-+} = 1 \, .
\end{eqnarray}
Explicit expressions for the Killing spinors corresponding to these superconformal supersymmetries will be obtained  in \S\ref{Killing} and will be required for localization in \S\ref{Solution}.
 
It is easy to see from the algebra that the generator
$Q = G^{++}_{+} + G^{--}_{-}$ squares to $4(L-J)$. Since $L$ is the generator of rotations of the Poincar\'e disk 
and $J$ is the generator of rotations of $S^{2}$, the square $Q^{2}$ is the generator of a compact bosonic 
symmetry. This is the generator that we will use for localization.

\section{Off-shell formulation of the theory \label{off-shell}}

In this section, we review the off-shell formulation of supergravity due to 
\cite{deWit:1979ug, deWit:1984px, deWit:1980tn}. 
This formalism has several  attractive features.
\begin{enumerate}
\item First, it allows the supersymmetry transformations 
to be realized in an off-shell manner which will be crucial for us to apply localization 
to the functional integral for quantum entropy.
\item Second, one can also include within 
the formalism a class of curvature squared corrections to the theory that are 
encoded in the Weyl multiplet. This has made  it possible to study the  higher derivative 
corrections to supersymmetric black holes using the full power of supersymmetry for solving 
BPS equations in the classical theory.  
\item Third, in the off-shell formalism, the supersymmetry transformations are specified  once and for all and do not need to be modified as one modifies the action with higher derivative terms. This is analogous to the situation for diffeomorphisms where the transformation properties of the metric, for example, are specified once and for all and does not depend on the form the action. Since the localization action that we use is constructed using these supersymmetry transformations, the localizing solutions that we will obtain by minimizing this action will therefore be universal and not dependent on the form of the physical action. This is clearly greatly advantageous both at the technical and conceptual level. 
\end{enumerate}

In this section we  rederive the classical properties of the black hole in this new language. 
This section is meant to set the stage and fix all the notations for the quantum calculation which 
we discuss in \S\ref{Solution}. It will therefore be concise; a detailed account of the off-shell 
formalism can be found,  for example, in the review \cite{Mohaupt:2000mj}. 

We  use the {\it conformal supergravity} approach to $\CN=2$ off-shell supergravity in four dimensions
 developed using {\it superconformal multiplet calculus}. 
The main idea is to extend the symmetries of the $\CN=2$ Poincar\'e supergravity
 to the $\CN=2$ superconformal algebra. This bigger algebra has dilatations, special 
conformal transformations, conformal $S$-supersymmetries, and  local $SU(2)' \times U(1)$ symmetries 
as extra symmetries compared to the Poincar\'e group\footnote{Note that the extra superconformal 
symmetry of this formalism is a gauge symmetry, not to be confused with the physical superconformal 
algebra of the near-horizon geometry of extremal black holes discussed in \S\ref{Supersymmetries} which is generated by the Killing vectors and Killing spinors of the  background.} . The conformal supergravity is 
then constructed as a gauge theory of this extended symmetry group. 

Upon gauge fixing the extra superconformal symmetries, one gets the Poincar\'e supergravity. 
In this sense, they are both gauge equivalent. However, the multiplet structure of the superconformal 
supergravity is smaller and simpler than the Poincar\'e theory. The form of the supersymmetry transformation 
rules is also simpler in the superconformal formalism, and one has a systematic way of deriving 
invariant Lagrangians. Following this approach, one gets an off-shell formulation of supergravity 
coupled to vector multiplets.

In \S\ref{Multiplets}, we first list the multiplets of the superconformal theory that will enter the theories we consider. 
In appendix \S\ref{susyvar}, we summarize some relevant aspects of the superconformal 
multiplet calculus including the supersymmetry variations of the various multiplets listed below. 
In \S\ref{Superaction} we discuss the invariant action of our interest.

\subsection{Superconformal multiplets \label{Multiplets}}

Our \emph{conventions} are as follows. In the Minkowski theory, all  fermion fields below are 
represented by Majorana spinors. In the Euclidean theory, they will be symplectic-Majorana. 
Greek indices $\mu,\nu,\dots$ indicate the curved spacetime, latin indices $a,b,\dots$ indicate 
the flat tangent space indices, and $i,j, \dots$ denote the $SU(2)'$ index. The $SU(2)'$ indices
are raised and lowered  by complex conjugation. 
$A^{-} \equiv \varepsilon_{ij}\,A^{ij}$ for any $SU(2)'$ tensor $A^{ij}$. We will also use
the superscript $\pm$ to denote (anti) self-duality in spacetime, the conventions should be clear from 
the context. 
We  use the  covariant derivative $D_{a}$, which  is defined to be covariant with respect to all the 
superconformal transformations as well as gauge fields of the theory if present.
The bosonic covariant derivative $\nabla_{a}$ is defined to be covariant with respect to all the 
bosonic transformations and the gauge fields, except the special conformal transformation.

We now summarize the field content of various multiplets.

\begin{enumerate}
\item {\it Weyl multiplet}: This is the gravity multiplet which contains all gauge fields arising from gauging the full superconformal symmetries. The field content is:
\be\label{Weylfields}
{\bf w} = \left( e_{\mu}^{a}, w_{\mu}^{ab}, \psi_{\mu}^{i}, \phi_{\mu}^{i}, b_{\mu}, f_{\mu}^{a}, A_{\mu}, \CV_{\mu \, j}^{\, i},  T_{ab}^{ij}, \chi^{i}, D \right) \, .
\ee
The fields $(e_{\mu}^{a}, w_{\mu}^{ab})$ are the gauge fields for translations (vielbien) and Lorentz transformations;
$\psi_{\mu}^{i}, \phi_{\mu}^{i}$ are the gauge fields for Q-supersymmetries and the  conformal S-supersymmetries; 
$(b_{\mu}, f_{\mu}^{a})$ are the gauge fields for dilatations and the special conformal transformations; and 
$(\CV_{\mu \, j}^{\, i}, A_{\mu})$ are the gauge fields for the $SU(2)'$ and $U(1)$ R-symmetries. Imposition of the `conventional constraints' determines $w_{\mu}^{ab}, \phi_{\mu}^{i},  f_{\mu}^{a}$ in terms of other fields and one is left with $24+24$ independent degrees of freedom. 
The $SU(2)'$ doublet of Majorana spinors $\chi^{i}$, the antisymmetric anti self-dual auxiliary field $T_{ab}^{ij}$ 
and the real scalar field $D$ are all auxiliary fields, some of which will play a non-trivial role later. 
This multiplet contains the gravitational degrees of freedom. 

\item {\it Vector multiplet}: The field content is
\be\label{Vectorfields}
{\bf X}^{I} = \left( X^{I}, \O_{i}^{I}, A_{\mu}^{I}, Y^{I}_{ij}  \right)
\ee
with $8+8$ degrees of freedom. $X^{I}$ is a complex scalar, the gaugini $\O^{I}_{i}$ are an $SU(2)'$ 
doublet of chiral fermions, $A^{I}_{\mu}$ is a vector field, and $Y^{I}_{ij}$ are an $SU(2)'$ triplet of 
auxiliary scalars. 
This multiplet contains the gauge field degrees of freedom. 

\item {\it Chiral multiplet}: The field content is
\be\label{Chiralfields}
{\bf \widehat{A}} = \left( \widehat{A},  \widehat{\Psi}_{i},  \widehat{B}_{ij},  \widehat{F}^{-}_{ab},  \widehat{\Lambda}_{i}, \widehat{C}  \right)
\ee
with $16+16$ components. $ \widehat{A},  \widehat{C}$ are complex scalars, $\widehat{B}_{ij}$ is a complex 
$SU(2)'$ triplet, $\widehat{F}^{-}_{ab}$ is an antiselfdual Lorentz tensor, and $ \widehat{\Psi}_{i},  \widehat{\Lambda}_{i}$ 
are $SU(2)'$ doublets of left-handed fermions. 
The action will also contain the conjugated right handed multiplet. 
One can impose a supersymmetric constraint on the chiral multiplet to get a reduced chiral 
multiplet with $8+8$ degrees of freedom. 

The covariant quantities of a vector multiplet 
are associated with a reduced chiral multiplet. 
The covariant quantities of the Weyl multiplet are also 
associated with a reduced chiral multiplet ${\bf W}^{ij}_{ab}$. Products of chiral multiplets 
are also chiral, and one thus gets a chiral multiplet 
${\bf \widehat{A}} = {\bf W}^{2}  = \ve_{ik} \ve_{jl} {\bf W}_{ab}^{ij} {\bf W}^{abkl}$. 
The lowest component of ${\bf \widehat{A}}$  is $\widehat A = (T^{ij}_{ab} \, \ve_{ij})^2$ and 
the highest component of ${\bf \widehat{A}}$ contains terms quadratic and linear in the curvature. 
The problem of building Lagrangians with terms quadratic in the curvature thus reduces to the 
simpler problem of coupling the chiral multiplet ${\bf \widehat{A}}$ to the superconformal theory.

\item {\it Compensating multiplet}: 
This multiplet will be used as a compensator to fix the extra gauge transformations.  
There are three types of compensators that have been used in the literature so far, a non-linear 
multiplet, a compensating hypermultiplet and a tensor multiplet. As an example, we 
discuss the non-linear multiplet \cite{Mohaupt:2000mj, Sahoo:2006rp}. 
Other multiplets have their relative advantages, in particular the compensating hypermultiplet 
is used extensively for the treatment of higher derivative terms \cite{LopesCardoso:2000qm}.\\
\ndt {\it Non-linear multiplet}:
\be\label{nonlinear}
\left( \Phi^{i}_{\,\a}, \lambda^{i}, M^{ij}, V_{a} \right)
\ee
where $\lambda^{i}$ is a $SU(2)'$ doublet spinor, $M^{ij}$ is a complex antisymmetric matrix of Lorentz scalars, 
and $V_{a}$ is a real Lorentz vector. $\Phi^{i}_{\, \a}$ is an $SU(2)'$ matrix of scalar fields with the $\a$ index 
transforming in the fundamental of a rigid $SU(2)'$, it describes three real scalars.  
Naively, the multiplet has $9+8$ degrees of freedom, but there is a supersymmetric constraint 
on the vector $V_{a}$ which reduces the degrees of freedom to $8+8$: 
\be\label{susycons}
D^a V_a - 3 D - \half V^a V_a - \frac{1}{4} |M_{ij}|^2 + D^a \Phi^i_{\;\;\alpha}
D_a \Phi^{\alpha}_{\;\;i} + \mbox{fermions} = 0
\ee

\end{enumerate}

\subsection{Superconformal action \label{Superaction}}

The procedure to get invariant actions is as follows: one first finds an invariant Lagrangian for a 
chiral multiplet, this was solved in \cite{deRoo:1980mm}. The second step is to 
write down a scalar function, the {\it prepotential} $F(X^{I})$ of the vector multiplets which is a 
meromorphic  homogeneous function of weight 2. One then uses the 
chiral Lagrangian of the first step for the chiral multiplet $\bf F$. This gives the 
two derivative $\CN=2$ Poincar\'e supergravity after gauge fixing. 
To include coupling to curvature square terms, one extends the function $F$ to depend on the lowest component
$\widehat A$ of the chiral multiplet $ \widehat{\textbf{A}} = {\textbf W^{2}}$.
$F(X^{I}, \widehat{{A}})$ is holomorphic and homogeneous of degree two in all its variables. 
One then uses the chiral Lagrangian of the first step for the chiral multiplet~$\bf F$. 

We  use the following notations. The prepotential which is a meromorphic function of its arguments 
obeys the homogeneity condition:
\be \label{homogen}
F(\lambda X, \lambda^2 \wh A) = \lambda^2
F( X, \wh A)\, .
\ee
Its various derivatives are defined as:
\be \label{defFI}
F_I = \frac{\p F}{ \p X^I}, \quad F_{\wh A} = \frac{\p F}{\p
\wh A}, \quad F_{IJ} =\frac{\p^2 F}{ \p X^I \p X^J},
\quad F_{\wh A I} = \frac{\p^2 F}{ \p X^I \p \wh A},
\quad F_{\wh A \wh A} =
\frac{\p^2 F}{ \p \wh A^2}\, .
\ee

Following the above procedure, one gets a  invariant action for $I=1,2,\dots , N_{V}+1$ vectors coupled 
to conformal  supergravity. The bosonic part of the action is:
\bea\label{supconaction}
e^{-1} {\cal L} & = & i \Big[ \bar{F}_I X^I ( \frac{1}{6} R - D )
+  {\nabla}_{\mu} F_I {\nabla}^{\mu} \bar{X}^I  \nonumber \\
& &  \quad  + \frac{1}{4}   F_{IJ} ( F^{-I}_{ab} - \frac{1}{4}  \bar{X}^I T^{ij}_{ab} \, \ve_{ij})
 ( F^{-abJ} - \frac{1}{4}  \bar{X}^J T^{ij}_{ab} \, \ve_{ij}) 
- \frac{1}{8}  F_I ( F^{+I}_{ab} - \frac{1}{4}  X^I T_{abij} \, \ve^{ij}) \, T^{ij}_{ab} \, \ve_{ij} \nonumber \\
 & & \quad - \frac{1}{8}   F_{IJ} Y^I_{ij} Y^{J ij}    - \frac{1}{32}  F \, (T_{abij} \, \ve^{ij})^{2} \nonumber \\
 & & \quad + \frac{1}{2}   F_{\widehat{A}} \widehat{C} - \frac{1}{8}   F_{\widehat{A} \widehat{A}}
(\ve^{ik} \ve^{jl} \widehat{B}_{ij} \widehat{B}_{kl} - 2 \widehat{F}^-_{ab} \widehat{F}^-_{ab}) 
+ \frac{1}{2}   \widehat{F}^{-ab} F_{\widehat{A}I} ( F^{-I}_{ab} - \frac{1}{4}  \bar{X}^I 
T^{ij}_{ab} \, \ve_{ij} )  \nonumber\\
 & &  \quad - \frac{1}{4}   \widehat{B}_{ij} F_{\widehat{A}I} Y^{Iij} \Big] + \rm{h.c.} 
\eea

To get to the $\CN=2$ Poincar\'e supergravity, one has to gauge fix the extra gauge transformations 
of the superconformal theory. To gauge fix the special conformal transformations, one sets the $K$-{\it gauge}:
\be\label{Kgauge}
b_{\mu} = 0 \ . 
\ee
To gauge fix the dilatations, one impose the $D$-{\it gauge}:
\be\label{Dgauge}
- i \big( X^{I} \bar{F}_{I}  - F_{I} \bar{X}^{I} \big) = 1 \ . 
\ee
To fix the chiral $U(1)$ symmetry, one fixes the $A$-{\it gauge}:
\be\label{Agauge}
X^{0} = \bar{X}^{0} \ . 
\ee
Due to these constraints on the scalars, the Poincar\'e supergravity has only $N_{V}$ independent scalars. 

In order to fix the $S$-supersymmetry, one imposes another gauge called the $S$-{\it gauge}. This constraint 
can be solved by eliminating one of the vector multiplet fermions. This gauge also breaks $Q$-supersymmetry, 
but a combination of the $S$ and $Q$ supersymmetries is preserved and corresponds to the physical 
supertransformations in the Poincar\'e theory. 

Finally, to fix the local $SU(2)'$ symmetry, one imposes the $V$-{\it gauge}:
\be\label{Vgauge}
\Phi^{i}_{\, \a} = \delta^{i}_{\, \a}
\ee

At each step in the gauge fixing process, one has to be careful to respect the previous gauge choices, and this 
leads to compensating field dependent transformations in the rules for the various remaining transformations. 
This is one of the reasons the final theory is more complicated. Finally, one has to solve algebraic 
equations to get rid of the auxiliary fields $D$ and $\chi$. 
At the end of this procedure, one gets the $\CN=2$ Poincar\'e supergravity with a bosonic Lagrangian:
\bea\label{PoincaresugraLag}
8 \pi e^{-1} {\cal L} &=& ( - i (  X^I\bar{F}_I - F_I \bar{X}^I )) \cdot  ( - \frac{1}{2} R ) \nonumber \\
& & + \big[ i  {\nabla}_{\mu} F_I {\nabla}^{\mu} \bar{X}^I  \nonumber \\
 & & +  \frac{1}{4} i F_{IJ} ( F^{-I}_{ab} -  \frac{1}{4} \bar{X}^I T^{ij}_{ab} \, \ve_{ij}) 
  ( F^{-abJ} -  \frac{1}{4} \bar{X}^J T^{ij}_{ab} \, \ve_{ij})
- \frac{1}{8} i F_I ( F^{+I}_{ab} - \frac{1}{4} X^I  T_{abij} \, \ve^{ij}) T^{ij}_{ab} \, \ve_{ij} \nonumber \\
 & & - \frac{1}{8} i F_{IJ} Y^I_{ij} Y^{J ij}  -\frac{i}{32} F \, (T_{abij} \, \ve^{ij})^{2} \nonumber \\
 & & + \frac{1}{2} i F_{\widehat{A}} \widehat{C} - \frac{1}{8} i F_{\widehat{A} \widehat{A}}
(\ve^{ik} \ve^{jl} \widehat{B}_{ij} \widehat{B}_{kl} - 2 \widehat{F}^-_{ab} \widehat{F}^-_{ab}) 
+ \frac{1}{2} i \widehat{F}^{-ab} F_{\widehat{A}I} ( F^{-I}_{ab} - \frac{1}{4} \bar{X}^I T^{ij}_{ab} \, \ve_{ij})  \nonumber\\
 & &  - \frac{1}{4} i \widehat{B}_{ij} F_{\widehat{A}I} Y^{Iij} + \rm{h.c.} \big]  \nonumber\\
 & & -i (  X^I\bar{F}_I - F_I \bar{X}^I ) \cdot ({\nabla}^a V_a 
- \frac{1}{2} V^a V_a - \frac{1}{4} | M_{ij} |^2 + D^a \Phi^i_{\;\;\alpha} D_a \Phi^{\alpha}_{\;\;i}) \;. \\ 
\nonumber
\eea
Note that both the covariant derivatives defined above are used in this expression, they are related by 
\be\label{defDa}
D^a V_a = {\nabla}^a V_a - 2 f^a_a + \mbox{fermionic terms} \;.
\ee


\section{Localization \label{Solution}}

We now turn to the  evaluation of  the supersymmetric black hole functional integral defined  in \S\ref{QuantumEntropy} using the localization techniques discussed in 
\S\ref{Localization}.  We use the  formalism of \S\ref{off-shell} so that the supercharge used for localization is realized off-shell.

The on-shell equations of motion that follow from the above Lagrangian \eqref{PoincaresugraLag} 
admit a half-BPS black hole solution 
\cite{LopesCardoso:1998wt,LopesCardoso:1999cv,LopesCardoso:1999xn,LopesCardoso:2000qm}. 
The near horizon geometry is an $AdS_{2} \times S^{2}$ which admits eight 
conformal supersymmetries\footnote{As mentioned above, these conformal supersymmetries 
are not the conformal supersymmetries of the four-dimensional theory discussed in the last section, 
the latter are gauge symmetries in that formalism.}.  
The values of other fields are determined by the attractor mechanism 
\cite{Ferrara:1995ih,Ferrara:1995h,Strominger:1996kf} in terms of the charges consistent with the 
isometries.  The near-horizon $AdS_{2} \times S^{2}$ geometry with the attractor values of the 
other fields can also directly be derived from the BPS equations \cite{Cardoso:2006xz}.

We first review  this  on-shell solution in \S{\ref{Onshell}}. We then proceed to find the 
localizing instanton solution in  \S{\ref{Offshell}} and evaluate the renormalized action for this solution 
in \S{\ref{RenAction}}.  We will sometimes refer to the localizing solution as the off-shell solution since 
for this solution the scalar fields are excited away from the attractor values inside the $AdS_{2}$. In \S\ref{What} we  put together  these ingredients to reduce the functional integral of $\hat W(q, p)$  to an ordinary integral on the localizing submanifold.

\subsection{On-shell attractor geometry \label{Onshell}}

Symmetries of $AdS_{2} \times S^{2}$ imply that various field in the near horizon region take the form
\bea \label{sol1}
&& ds^2=  v \left[-(r^2-1)dt^2+\frac{dr^2}{r^2-1}\right] + v  \, \left[ d\psi^2 + \sin^2 (\psi) d \phi^2 \right] \ , 
 \nonumber \\
&& F^I_{rt} = e^{I}_{*}, \quad F^I_{\psi\phi}=  p^I
\, \sin\psi, \quad 
X^I = X_{*}^I, \quad
T^-_{rt}= v \, w \ , \nonumber \\
&& D-\frac{1}{3} R = 0, 
\quad M_{ij}=0, \quad \Phi^\alpha_i
=\delta^\alpha_i\, , \quad Y^I_{ij}=0 \ . 
\eea 
The values of the constants $(e^{I}_{*}, X^{I}_{*},v_{*})$ that appear in this solution are determined 
in terms of the charges $(q_{I}, p^{I})$ 
by the attractor equations which follow from the BPS conditions \cite{LopesCardoso:1998wt},
 or, equivalently using the entropy function formalism \cite{Sahoo:2006rp}:
\bea 
 && v  = \frac{16}{ \bar w w} \, , \quad  \hat A =  -4 \omega^{2} \ ,  \label{sol2a}  \\
 && e^I_{*} - i  p^I - \frac{1}{2} \bar X_{*}^I v w =  0 \ ,\label{sol2b}  \\ 
 && 4 i (\bar w^{-1}\bar F_I - w^{-1} F_I) =  \, q_I \ . \label{sol2c}  
\eea
Taking the real and imaginary parts of \eqref{sol2b} and substituting \eqref{sol2a} gives
\bea \label{scalarattval}
 4 (\bar w^{-1}\bar X_{*}^I + w^{-1} X_{*}^I)&=& e^I_{*} \, ,  \label{sol3a}
\\
4 i (\bar w^{-1}\bar X_{*}^I - w^{-1} X_{*}^I) &=&  p^I \, , \label{sol3b}
\ee 
where $F_{I}$  should be thought of as functions of $X^{I}_{*}$.
This geometry preserves eight superconformal supersymmetries as reviewed \S{\ref{Killing}} which   extends the symmetries to the supergroup $SU(1, 1|2) \rtimes SU(2)'$ discussed in \S{\ref{Supersymmetries}}. The field $w$ can be fixed by a gauge choice.
In the rest of the paper, we choose a gauge in which $w = \bar w = 4$ using the  local scaling symmetry of the Lagrangian and the $U(1)$ invariance. In this gauge, the radius $v$ of  both $AdS_{2}$ and $S^{2}$ 
equals one, this simplifies the discussion of Killing 
spinors\footnote{This is different from  the gauge used in the previous section and also from the gauge $\omega = 8$ which is commonly used. 
These gauge choices do not affect considerations in this paper, but a  better understanding of different gauge choices can be useful to simplify the analysis. We plan to return to this issue in  future. 
}.

\subsection{Localizing action and the localizing instantons \label{Offshell}}

In order to use the technique of localization for our system, we need to pick a subalgebra of the 
full supersymmetry algebra discussed in  \S\ref{Supersymmetries}, whose bosonic generator is compact. 
We shall choose the subalgebra generated by the action of the supercharge 
\be\label{defQ1}
Q_{1} = G^{++}_{+} + G^{--}_{-} \ , 
\ee
which generates the compact $U(1)$ action:
\be\label{Q1anticom}
Q_{1}^{2} = 4 ( L - J ) \ . 
\ee
The  explicit form of the Killing spinors can be found in 
\S{\ref{Killing}}. The above choice of the supercharge corresponds to choosing the supersymmetry parameter $\zeta_{1}$ defined in \eqref{defzeta}.
In this section, we  use the notation $Q \equiv Q_{1}$, $\zeta \equiv \zeta_{1}$. 

The localizing Lagrangian is then defined by 
\begin{equation}\label{loclag}
\CL^{Q} := QV \quad {\rm with} \quad V := (Q \Psi, \Psi) \, ,
\end{equation}
where $\Psi$ refers to all fermions in the theory. The localizing action is then defined by
\begin{equation}
S^{Q} = \int d^{4} x \sqrt{ g} \, \CL^{Q} \, .
\end{equation}
The localization equations that follow from this action  are
\begin{equation}
Q \Psi = 0 \, . 
\end{equation}
These are the equations that we would like to solve.

We assume that the supergroup isometries of the near horizon geometry are not broken further by the Weyl multiplet fields. 
By construction, as long as these symmetries are maintained, the fermions of the Weyl multiplet do not transform under the action of  $Q$
 \eqref{Weylvar1} --\eqref{Weylvar3} in the $AdS_{2}$ attractor background. 
One can check that the fermions of the chiral multiplet and the non-linear multiplet also do not transform 
in this background. This prompts us to look for solutions where one still has the $AdS_{2}$ attractor geometry, but the scalars of the 
vector multiplets can move away from their attractor values\footnote{Solutions more general than our simplifying ansatz are in principle possible where the Weyl multiplet fields also vary inside the $AdS_{2}$ .}. As we will see there do exist nontrivial solutions where the vector multiplet fields get excited maintaining the symmetries of the attractor geometry.

The action of $Q$ on the fermionic field of the vector multiplet takes the form
\begin{eqnarray}
 && Q \, \Omega^{Ii}_{+}=\frac{1}{2}(F_{\mu\nu}^{I-}-\frac{1}{4}\bar{X}^{I} \, T^{-}_{\mu\nu}) \, 
 \gamma^{\mu} \, \gamma^{\nu} \, \zeta^{i}_+ +2i \displaystyle{\not}\partial X^{I} \, \zeta^i_-+Y^{Ii}_j \, \zeta^j_+ \ , \\
 && Q \, \Omega^{Ii}_-=\frac{1}{2}(F_{\mu\nu}^{I+}-\frac{1}{4}X^{I} \, T^{+}_{\mu\nu}) \, 
 \gamma^{\mu} \, \gamma^{\nu} \, \zeta^{i}_- +2 i \displaystyle{\not}\partial \bar{X}^{I} \, \zeta^i_+ +Y^{Ii}_j \, \zeta^j_- \ .
\end{eqnarray}

Let us recall the attractor equations for the constant values of the various fields 
in terms of the  electric gauge field strengths $e^{I}$ and the magnetic charges $p^{I}$:
\begin{eqnarray} \label{attractor eqs again}
&& e_{*}^{I} - ip^{I} - 2\bar{X}_{*}^{I} = 0 \ ,  \qquad e_{*}^{I} + ip^{I} - 2 X^{I}_{*} = 0 \, \qquad Y^{I}_{ij*} = 0 \  .
\end{eqnarray}
We are interested in the off-shell solutions in which the vector multiplet scalars $X^{I}$ move away from their attractor values $X^{I}_{*}$. We therefore parametrize the off-shell $X^{I}$ fields as
\begin{equation}\label{offshellpar}
X^{I} := X_{*}^{I } + \Sigma^{I}\, , \qquad \bar X^{I} := \bar X_{*}^{I } + \bar \Sigma^{I},
\end{equation}
 so that $\Sigma^{I}$ and $\bar \Sigma^{I}$ are values the scalar fields away from the attractor values. We proceed analogously for the gauge and auxiliary fields
 \begin{equation}\label{offshellpar3}
  F_{\mu\nu}=F^{*}_{\mu\nu}+f_{\mu\nu},\; Y^i_j=K^i_j
 \end{equation}
 The attractor background (\ref{attractor eqs again}) is a solution of $Q\Omega=0$ and therefore the attractor values drop out from the supersymmetry variations. This means
\begin{eqnarray}\label{Qvars}
 &&Q\Omega^{Ii}_{+}=\frac{1}{2}(f^{I-}_{ab}-\frac{1}{4}\bar{\Sigma}^I T^{-}_{ab})\gamma^{a}\gamma^{b}\xi^i_{+}+2 i\slashed \partial \Sigma^I \xi^i_{-}+K^{Ii}_{j}\xi^{j}_{+}\\
&&Q\Omega^{Ii}_{-}=\frac{1}{2}(f^{I+}_{ab}-\frac{1}{4}\Sigma^I T^{+}_{ab})\gamma^{a}\gamma^{b}\xi^i_{-}+2 i\slashed \partial \bar{\Sigma}^I \xi^i_{+}+K^{Ii}_{j}\xi^{j}_{-}
\end{eqnarray}where $a,b$ are the Euclidean tangent space indices. 
In Euclidean supersymmetric theories, the R-symmetry is $SU(2)'\times SO(1,1)$. As a consequence both $\Sigma,\bar{\Sigma}$ and $f_{ab}$ are real\footnote{The reality conditions 
contained an error in the previous version of the paper which was pointed out in \cite{Gupta:2012cy}, 
where a complete analysis of the Weyl multiplet was carried out. 
We present the correct analysis here. The final solutions remain the same as in the previous version.}  \cite{Cortes:2003zd}. 
We then consider adapted coordinates defined in the following way:
\begin{equation}
 \Sigma=H-J;\;\;\bar{\Sigma}=H+J
\end{equation}
Note that  $Y^{I}_{ij}$ are triplets under the  $SU(2)'$ rotation. We assume 
that for the BPS equations that we solve, they all have to be aligned along the same direction 
in the $SU(2)'$ space\footnote{This assumption was later justified in \cite{Gupta:2012cy}.}. 
Hence  we parametrize them as 
\begin{eqnarray}\label{offshellpar2}
  Y^{I1}_{\,\,\, 1}=-Y^{I2}_{\,\,\,2}=K^{I} \, ; \qquad 
 Y^{I1}_{\,\,\,2}=  Y^{I2}_{\,\,\,1} = 0   \, .
\end{eqnarray}
With this parametrization, we can add the two equations  \eqref{Qvars} and perform a Euclidean 
continuation to obtain
\begin{equation}
 Q\Omega^{Ii}=\frac{1}{2}f_{ab}\gamma^{a}\gamma^{b}\xi^i+2i\slashed \partial H^I \xi^i+ 2i\slashed \partial J \gamma_5 \xi^i-2i H^I \gamma^{0}\gamma^{1}\xi^i-2i J^I \gamma^{2}\gamma^{3}\xi^i+K^i_j \xi^j
\end{equation}with $\Omega^i=\Omega^i_+ + \Omega^i_{-}$ and $\xi^i=\xi^i_+ + \xi^i_{-}$.
Note that the $a, b$ are tangent space indices and all gamma matrices $\gamma^{a}$ above are constant matrices of Euclidean $\mathbb{R}^{4}$.

The inner product for spinors $\chi_{1}$ and $\chi_{2}$ in Euclidean space is simply 
\begin{equation}\label{inner}
(\chi_{1} \, , \chi_{2}) = \chi_{1}^{\dagger} \chi_{2} \, .
\end{equation}
With this inner product, the  localization Lagrangian \eqref{loclag} restricted to only the vector multiplet fermions is given by
\begin{equation} 
\CL^{Q} =  QV  : = Q(Q\Psi, \Psi)
\end{equation} 
with $V$ chosen as in \eqref{loclag} with $\Psi$ denoting the vector multiplet fermions. Note that $V$ is H-invariant because $\zeta$ is independent of the combination $\theta -\phi$ and $ H$ is the vector field that generates translations along $\theta -\phi$. 
The bosonic part of this Lagrangian is
\begin{equation}
\CL_{\rm bos}^{Q} \equiv QV \big|_{\text{bosonic}}= \sum_{I=0}^{n_{V}}  (Q \Omega^{I}, {Q\Omega^{I}}) \ .
\end{equation}
With our choice of the inner product \eqref{inner} this Lagrangian is manifestly positive definite producing a well defined $\lambda\rightarrow \infty$ limit in (\ref{exact deform}).

The choice of $Q$ is determined by the choice of the Killing spinor $\zeta$.  Substituting the explicit form of the Killing spinor $\zeta$ and the gamma matrices defined in \S{\ref{Killing}},   the bosonic Lagrangian $ \mathcal{L}_{\rm bos}^{Q}$ as a function of the fields $H, J, K, f$ can be evaluated after a somewhat tedious algebra. 
We find that $\frac{1}{2}\CL^{Q}_{\rm bos}$ equals
\begin{eqnarray}\label{LQdef}
 S_{loc}&=&\sum_{a,b}\frac{\alpha}{8}(f_{ab}^+)^2+\sum_{a,b}\frac{\beta}{8}(f_{ab}^-)^2 \nonumber\\
&+& 2\alpha \left[(\partial_0\Phi^+)^2+(\partial_2\Phi^+)^2\right]+2\beta \left[(\partial_0\Phi^-)^2+(\partial_2\Phi^-)^2\right]\nonumber\\
&+& 2\alpha \left[\Phi^+ +\frac{1}{\alpha}\left(\partial_1 \Phi^-\sinh(\eta)-\partial_3\Phi^- \sin(\psi)-\frac{K}{2}(1+\cos(\psi)\cosh(\eta)) \right)\right]^2 \nonumber\\
&+& 2\beta \left[\Phi^- +\frac{1}{\beta}\left(\partial_1 \Phi^+\sinh(\eta)+\partial_3\Phi^+ \sin(\psi)-\frac{K}{2}(1-\cos(\psi)\cosh(\eta)) \right)\right]^2 \nonumber\\
&+&\frac{2}{\alpha}\left[\partial_1 \Phi^-\sin(\psi)+\partial_3 \Phi^- \sinh(\eta)+\frac{K}{2}\sin(\psi)\sinh(\eta)\right]^2 \nonumber\\
&+&\frac{2}{\beta}\left[-\partial_1 \Phi^+\sin(\psi)+\partial_3 \Phi^+ \sinh(\eta)-\frac{K}{2}\sin(\psi)\sinh(\eta)\right]^2
\end{eqnarray}where we have defined $\Phi^+=H+J$ and $\Phi^-=H-J$, $\partial_a=e^{\mu}_a\partial_{\mu}$ and $\alpha=\cosh(\eta)+\cos(\psi)$,  $\beta=\cosh(\eta)-\cos(\psi)$. 

It is understood that in (\ref{LQdef}) all squares are summed over the index $I$. 
Recall that $a = 0, 1, 2, 3$ correspond to the directions along the coordinates $\theta, \eta, \phi, \psi$ respectively used for example in \eqref{metric2}.
Since $\alpha$ and $\beta$ are non-negative,  $\CL^{Q}_{\rm bos}$ is a sum of positive squares. 

The minimization equations now follow by setting each of the squares in \eqref{LQdef} to zero. This leads to first order differential equations for various fields which have to be solved with boundary conditions  consistent with the definition of the original functional integral on  Euclidean $AdS_{2}$ space. Equations \eqref{offshellpar},  \eqref{offshellpar2}, \eqref{offshellpar3} imply that fields $\Sigma^{I}$ and $K^{I}$ and $f^{I}$  must vanish at the boundary. 

Part of the equations readily imply that $f_{ab}=0$ and $\partial_{0,2}H=\partial_{0,2}J=0$. The remaining equations are
\begin{eqnarray}
 && \cosh(\eta)H+\cos(\psi)J+\partial_1 H\sinh(\eta)+\partial_3 J\sin(\psi)-\frac{K}{2}=0 \label{eq1}\\
&& \cosh(\eta)J+\cos(\psi)H-\partial_1 J\sinh(\eta)-\partial_3 H\sin(\psi)-\frac{K}{2}\cos(\psi)\cosh(\eta)=0\label{eq2}\\
&& \partial_1 H \sin(\psi)=\partial_3 J \sinh(\eta)-\frac{K}{2}\sin(\psi)\sinh(\eta)\label{eq3}\\
&& \partial_3 H \sinh(\eta)=\partial_1 J \sin(\psi)\label{eq4}
\end{eqnarray} which is a complicated system of coupled first order differential equations. After some analysis, that we refer to the appendix \ref{appB}, we find
\begin{equation}\label{CHCKsol}
 H^I=\frac{C^I}{\cosh(\eta)},\;\;K^I=\frac{2C^I}{\cosh(\eta)^2},\;\; J^I=0.
\end{equation}

We  have thus succeeded in finding a family of exact  solutions to the localization equations
which respect the classical boundary conditions on $AdS_2$ and are smooth everywhere in the interior. 
In terms of the original variables defined in \eqref{offshellpar}, we have 
\begin{eqnarray} \label{HEKEsol}
&& X^{I}  =  X^{I}_{*} +  \frac{C^{I}}{\cosh(\eta)} \ , \qquad  \bar X^{I}  =  \bar X^{I}_{*} +  \frac{C^{I}}{\cosh(\eta)}\\
&& \qquad \qquad Y^{I1}_{1} = - Y^{I2}_{2} =  \frac{2C^{I}}{\cosh(\eta)^2} \,\,  .
\end{eqnarray}
Since the scalar  fields are now
excited away from their attractor values, they are no longer at the minimum  of the classical entropy function.  Even though scalar fields `climb up' the potential away from the minimum of the entropy function the solution remains Q-supersymmetric (in the Euclidean theory) because an auxiliary field gets 
excited appropriately to satisfy the Killing spinor equations.

It is worth pointing out that the solutions \eqref{CHCKsol} look much simpler if we use the conformal transformation \eqref{conformal-trans} in \S\ref{Supersymmetries} to map $AdS_{2} \times S^{2}$ to $S^{4}$. Since the scalar fields $X$ and the auxiliary fields $Y$  have Weyl weight 1 and 2 respectively,
and since the conformal factor is $\cosh(\eta)$,  the fields $\Sigma$ and $Y$  are simply constant   on $S^{4}$. 
This is very similar to the localizing solution found by Pestun \cite{Pestun:2007rz} in a very different context of computing the expectation value of Wilson line in super Yang-Mills theory on $S^{4}$. Of course, under this conformal transformation the attractor values also will transform and since they are constant on $AdS_{2} \times S^{2}$, they  will no longer be constant on $S^{4}$. It is therefore more natural to work in the $AdS_{2} \times S^{2}$ frame. In any case, for computing the quantum entropy, the $AdS_{2}$ boundary conditions play an important role as we will see in the next subsection.    As pointed out in \cite{Banerjee:2009af}, in this frame our computation  has close formal similarity with the gauge theory computation of `t Hooft-Wilson line in the formulation of \cite{Kapustin:2005py, Gomis:2009ir} which could be useful in the computation of one-loop determinants and the instanton contributions. Note  that we are using localization techniques to evaluate a bulk functional integral of 
supergravity whereas in \cite{Pestun:2007rz, Kapustin:2005py, Gomis:2009ir} it was used to evaluate a functional integral in the boundary gauge theory.

\subsection{Renormalized action for the localizing instantons \label{RenAction}}

To obtain the exact macroscopic quantum partition function we would like to evaluate the renormalized action restricted to the  submanifold $\mathcal{M}_{Q}$ in field space of localizing instantons. We will find that even though both the original action and the solution are rather complicated, the renormalized action  is a remarkably simple function   of the collective coordinates $\{C^{I}\}$ determined entirely by the prepotential. Recall that the renormalized action defined in  \S\ref{QuantumEntropy} takes the form
\be\label{Sren2}
\CS_{\rm ren} :=   \CS_{\rm bulk}  + \CS_{\rm bdry} + i \, \frac{q_I }{2}  \int_{0}^{2\pi} A^I_{\theta}  \, d\theta \ .
\ee
The charges used here are related to the ones used in \eqref{Sren} by  $q_{I} =  -2 	q_{i}$ to be consistent with the normalization of gauge fields used in the literature, for example, in the reviews \cite{Sen:2007qy, Mohaupt:2000mj}.

We proceed to evaluate the bulk action given as a four dimensional integral of the  the supergravity Lagrangian \eqref{PoincaresugraLag} over $AdS_{2} \times S^{2}$. 
We note first that since various auxiliary  fields vanish for the off-shell solution, 
the Lagrangian  \eqref{PoincaresugraLag} simplifies to (recall $\widehat A = (T^{ij}_{ab} \, \ve_{ij})^2$):
\begin{eqnarray}
 8\pi \, \mathcal{L} & = & -\frac{i}{2} \, (X^I\bar{F}_I-\bar{X}^IF_I) \, R
  +\big[i \, \partial_{\mu}F_I \, \partial^{\mu}\bar{X}^I 
  +\frac{i}{4} \, F_{IJ} \, (F^{-I}_{\mu\nu}-\frac{1}{4}\bar{X}^I \, T^{ij}_{\mu\nu} \ve_{ij}) 
    (F^{-J\mu\nu}-\frac{1}{4}\bar{X}^J \, T^{\mu\nu ij} \ve_{ij}) \nonumber\\
&& \quad +\frac{i}{8} \, \bar{F}_I \, (F^{-I}_{\mu\nu}-\frac{1}{4}\bar{X}^I \, 
   T_{\mu\nu ij} \ve^{ij}) \,T^{ij}_{\mu\nu} \ve_{ij}
   -\frac{i}{8} \, F_{IJ} \, Y^I_{ij} \, Y^{Jij} + \frac{i}{32} \, \bar{F} \, \hat{A} 
   + \frac{i}{2} \, F_{\hat{A}} \, \hat{C} + {\rm h.c.} \big] \ . 
\end{eqnarray}
Moreover,  for  $AdS_2\times S^2$ both the Ricci scalar $R$ and the Weyl tensor 
$C$ are zero. 
Substituting  $X^I = X^{I}_{*} + \Sigma^{I}$  and $\bar X^I =\bar X^{I}_{*} + \bar \Sigma^{I}$ from \eqref{offshellpar} and using the attractor equation \eqref{attractor eqs again}
in the form 
\begin{equation}
F^{-I}_{\mu\nu}-\frac{1}{4}\bar{X_{*}}^I \, T^{ij}_{\mu\nu} \ve_{ij} = 0 \ , 
\end{equation}
we get 
\begin{eqnarray}
 8\pi \, \mathcal{L} = i \, F_{IJ} \, (\partial_{\eta}\bar{\Sigma}^I) (\partial_{\eta}\Sigma^J)
  -i \, F_{IJ} \bar{\Sigma}^I \, \Sigma^J +\frac{i}{4} \, F_{IJ} \, K^I  K^J 
  +2i \, \bar{F}_I \, \bar{\Sigma}^I- 2i \, \bar{F} + {\rm h.c.} \ . 
\end{eqnarray}
Substituting the solution \eqref{HEKEsol} into the above equation, we find that  the first three terms of add up  to zero.
We are thus left with 
\begin{equation} \label{Lstep}
 8\pi\mathcal{L}=2i\bar{F}_I\bar{\Sigma}^I-2i\bar{F}+ h.  c. \, .
\end{equation}
Since we keep the classical values $X^{I}_{*}, \bar X^{I}_{*}$ fixed in this problem, 
differentiating with respect to $X^{I}$ is the same as differentiating with respect to $\Sigma^{I}$.
This can be explicitly evaluated to find
\begin{eqnarray} \label{Lstep3}
 8\pi \, \mathcal{L}  =  2 i \p_{r} \left(r (F - \bar F) \right) \ , \quad \textrm{with} \quad \Sigma^{I} = \frac{C^{I}}{r} \, .
\end{eqnarray}
%
The $\mathcal{N}=2$ supergravity Euclidean action is
\begin{equation}
 \mathcal{S}_{\rm bulk} =  \int d^4x \sqrt{g} \,  \mathcal{L} \, .
\end{equation}
The off-shell fields do not depend on  the coordinates of the  $S^{2}$ and the angular variable $\theta$ of the $AdS_{2}$. These integrals can be done trivially and give an overall factor of $8\pi ^{2}$, so that 
\begin{eqnarray} \label{Sbulk}
 \CS_{\rm bulk}  & = &  {8 \pi^{ 2}} \int_0^{\eta_{0}} \CL \,  \sinh(\eta) \,  d\eta \  
 = 8\pi^{2} \int_1^{r_{0}} \CL \, dr \ , \\
& = &    2 \pi i \int_1^{r_{0}} dr \partial_{r} \left(r (F - \bar F) \right) \ , \\
& = &  2 \pi i r_{0} \Big [F\big(X_{*}^{I} + \frac{C^{I}}{r_{0}}\big) - \bar  F\big(X_{*}^{I} + \frac{C^{I}}{r_{0}}\big) \Big]  
 - 2 \pi i \Big[F(X_{*}^{I} + C^{I}) -  \bar F(X_{*}^{I} + C^{I}) \Big]. \label{Sbulkeval} 
\end{eqnarray}
The first piece in \eqref{Sbulkeval} which is linear in $r_{0}$ can be rewritten as:
\begin{eqnarray} \label{Sbulk1}
&& 2 \pi i r_{0} \Big(F\big(X_{*}^{I} + \frac{C^{I}}{r_{0}}\big)  - \bar F\big(X_{*}^{I} + \frac{C^{I}}{r_{0}}\big) \Big) =  \cr
& & \qquad \qquad =   2 \pi i r_{0} \big(F(X_{*}^{I}) -  \bar F(X_{*}^{I}) \big) + 2 \pi i (F_{I}(X_{*}^{I}) - \bar F_{I}(X_{*}^{I})) \, C^{I} + \CO(1/r_{0})\cr
 & &  \qquad \qquad =    2 \pi i r_{0} \big(F(X_{*}^{I}) -  \bar F(X_{*}^{I}) \big) - 2 \pi q_{I} C^{I} + \CO(1/r_{0}) \ , 
\end{eqnarray}
where we have used a Taylor expansion in the first line and the attractor equation 
\be
F_{I}(X_{*}^{I}) - \bar F_{I}(X_{*}^{I}) = i q_{I}
\ee
in the second. 

The Wilson line evaluates to 
\be\label{Wilsoneval}
  i \, \frac{q_I }{2}  \int_{0}^{2\pi} A^I_{\theta}  \, d\theta \  =  \pi q_{I} e^{I}_{*} (r_{0} -1) \ . 
\ee
Hence we choose 
\be\label{Sbdry}
\CS_{\rm bdry} =  - 2 \pi r_{0} \left(  \frac{q_{I} \, e^{I}_{*}}{2} + i \, \big(F(X_{*}^{I}) -  \bar F(X_{*}^{I}) \big)  \right) \ . 
\ee
so that $\CS_{ren} =  \CS_{\rm bulk} + \CS_{\rm bdry} + i \frac{q}{2} \oint A$ is finite. 

As reviewed in \S\ref{QuantumEntropy},  the main purpose 
of the boundary action is to cancel the divergence in the bare bulk action plus Wilson line
which grows linearly with the length of the boundary. 
In order to cancel this divergence, 
we use a boundary cosmological constant which 
must be specified along with the other boundary data. 
Indeed we have found that  $\CS_{\rm bdry}$ which is a constant that grows linearly 
with the length of the boundary indeed only depends on the fixed charges and not on the 
fluctuating fields. 

In general, however, there could be a finite part of the boundary action which does depend 
on the fields that are integrated over. The full boundary action should be constrained by 
supersymmetry. We shall discuss the supersymmetry of the functional integral 
in appendix \S\ref{Supersymmetry}. The conclusion of the analysis in appendix \S\ref{Supersymmetry} 
is quite simple -- the  finite part of the boundary action in our problem actually vanishes 
due to supersymmetry, and therefore the above prescription for $S_{\rm ren}$ as a sum of terms \eqref{Sbulk}, 
\eqref{Sbdry} and \eqref{Wilsoneval} is already supersymmetric. In appendix  \S\ref{Supersymmetry},
we shall rewrite the above in a manner that is manifestly supersymmetric. This 
rewriting takes the form of a functional integral with a supersymmetric Wilson line 
\cite{Maldacena:1998im, Rey:1998ik}  with the bulk action as above \eqref{Sbulk}, and a boundary action which exactly cancels the boundary piece in  \eqref{Sbulkeval}.

We thus  obtain the following expression for the renormalized action:
\be\label{Srenfinal}
\CS_{\rm ren} =  - \pi \, q_I \, e^I_* - 2 \pi  q_{I} C^{I}  
 - 2 \pi i \big(F(X_{*}^{I} + C^{I}) -  \bar F(X_{*}^{I} + C^{I}) \big) \ , 
\ee
The notation $e^{I}_{*}$ refers to the classical values of the electric field strengths as a function of the 
charges $(q_{I}, p^{I})$.
Using the scalar attractor values  \eqref{scalarattval}, 
and the new variable 
\be\label{ephi}
\phi^I \equiv e_{*}^I+2 C^I \ ,
\ee
we can express the renormalized action in a remarkably simple form:
\begin{eqnarray}
 \mathcal{S}_{ren}(\phi, q, p) =  - \pi q_I\phi^I + \mathcal{F}(\phi, p)\, .
\end{eqnarray}
with
\begin{equation} \label{freeenergy2}
\mathcal{F}(\phi, p) = - 2\pi i \left[ F\Big(\frac{\phi^I+ip^I}{2} \Big) -
 \bar{F} \Big(\frac{\phi^I- ip^I}{2} \Big) \right] \, .
 \end{equation}
Note that the electric field remains fixed at the attractor value but $\phi^{I}$ can still fluctuate with  $C^{I}$ taking values over the  real line. We will discuss the significance of this fact in 
\S\ref{Connection}. Note also that the prepotential is evaluated at precisely for values of the scalar fields at the origin of $AdS_{2}$ and not at the boundary of $AdS_{2}$.
Thus the classical contribution to the localization integrand will be of the form
\begin{equation}
e^{S_{ren}} =  e^{- \pi  \phi^{I} q_{I} + \mathcal{F}(\phi, p)}
\end{equation}
There will be additional contribution to the integral which we discuss next.

\subsection{Evaluation of $\hat W(q, p)$ \label{What}}

We have thus determined which field configuration to integrate over and the classical action for these configuration.
The full functional integral will require three additional ingredients.

\begin{myitemize}
\item The integration measure over the $\{C^{I} \}$ fields over 
the submanifold $\mathcal{M}_{Q}$ of critical points of $Q$ simply  descends from the measure $\mu$ of supergravity over the field space $\mathcal{M}$. We denote this measure by $[dC]_{\mu}$ which can be computed using standard methods of collective coordinate quantization.  
\item There will  be one-loop determinants of fluctuations around the localizing manifold which can be evaluated from the quadratic piece of the localizing action $S^{Q}$.  We denote  this determinant contribution by $Z_{det}$. It is in  principle a straightforward but technically involved computation. Very similar determinants have been analyzed in  detail  for gauge theory \cite{Pestun:2007rz}. In string theory, the one-loop determinants and the duality invariance  measure around the on-shell solution have been analyzed in \cite{Banerjee:2010qc} and in \cite{LopesCardoso:2006bg, Cardoso:2008fr} respectively.  Some aspects of these  computations both from gauge theory and from around the on-shell saddle point  could be adapted to study the measure and determinants around our off-shell instantons solutions \cite{Dabholkar:2010t}.  
\item In addition, there will be a contribution from point  instantons and anti-instantons viewed as singular configurations that couple to the vector multiplet fields as long as they preserve the same supersymmetry.  In gauge theory computations \cite{Pestun:2007rz}, the instantons will  be localized at the center of $AdS_{2}$ and at the north pole of the $S^{2}$ whereas the anti-instantons will be be localized at the center of $AdS_{2}$ and at the south pole of the $S^{2}$. Since string theory contains gauge theory at low energies we expect a similar structure also in string theory.  We denote this generating function for the instantons  by $Z_{inst}$. The generating function for anti-instantons will be the complex conjugate of the generating function for instantons. We will thus get a factor of $|Z_{inst}|^{2}$ which will  depend on the details of the string compactification, the spectrum of wrapped brane-instantons, and  the duality frame under consideration.
In gauge theory this generating function is the equivariant instanton partition function computed by Nekrasov \cite{Nekrasov:2002qd}. Since the low energy limit of string theory will reduce to gauge theory on $AdS_{2} \times S^{2}$, it would be interesting to explore if there are generalization of the gauge theory results to string theory. 
 \end{myitemize}

Putting these ingredients together we can conclude that the functional integral will have the form
\begin{equation}\label{integral2}
 \hat W(q, p) = \int_{\mathcal{M}_{Q}}  e^{-\pi  \phi^{I} q_{I}} e^{\mathcal{F}(\phi,  p)} |Z_{inst}|^{2} \, Z_{det} \, \, [dC]_{\mu}\,     .
\end{equation}
We have thus successfully reduced the functional integral to ordinary integrals. The dominant piece of the answer given by $e^{-S_{ren}}$ we have already evaluated explicitly.

In specific string compactifications the undetermined factors $Z_{det}$ and $|Z_{inst}|^{2}$ can simplify. For example, with $\CN=4$ supersymmetry, in gauge theory  both $|Z_{instanton}|^{2}$ and $Z_{det}$ equal unity.  Similarly, it was found in  \cite{Banerjee:2010qc} that very similar determinant factors for vector multiplets equal unity $\CN=4$ theories. One expects that this simplification will extend to the factors appearing in \eqref{integral2} around the localizing solution in $\CN=4$ theories.

\section{Quantum entropy and the topological string\label{Connection}}

We now turn  to the original problem of evaluating of $W(q, p)$.  There are several issues that have to be addressed to extend the supergravity computation to a full string computation.

\begin{myitemize}
\item First, the full action of string theory of course contains more fields in addition to vector multiplets, in particular the hyper multiplets. 
\item Second, even if we restrict our attention to vector multiplets, the action will in general contain not just the F-terms which are chiral superspace integrals but also the D-terms which are  nonchiral superspace integrals. 
\item Third, there can be additional contributions  from functional integral over  orbifolds of $AdS_{2}$ that are allowed in the full string theory but not visible in supergravity. 
\end{myitemize}
 We discuss these questions below.

\subsection{D-terms,  hyper-multiplets, and evaluation of $W_{0}(q, p)$\label{Eval}}

We have thus far considered only F-type terms for the action of the vector multiplets which are chiral integrals over $\mathcal{N} =2$ superspace of the form $\int d^{4}\theta$. 
The effective action of string theory will contain  in general  D-type terms which are nonchiral  integrals over $\mathcal{N} =2$ superspace of the form $\int d^{4}\theta d^{4}\bar \theta$.  It is not \textit{a priori} clear that these terms will not contribute to the functional integral. We would like to make  the following two observations in this connection.
\begin{myitemize}
\item Since our localizing action $S^{Q}$ follows from off-shell supersymmetry transformations,  it does not depend on what terms  are present in the physical action $S$.  Hence our localizing instanton solutions are universal and they will continue to exist even with the addition of the  D-terms. The question then reduces to evaluating  the D-terms on these solutions to obtain their contribution to the renormalized action.
\item It has recently been shown \cite{deWit:2010za} that a large class of D-type terms do not contribute to the Wald entropy.  This class of terms are constructed using  the `kinetic multiplet' $T$  obtained from a chiral multiplet  $\Phi$  of Weyl weight $0$  by $T = \bar D^{4} \bar \Phi$ which transforms like a chiral multiplet of Weyl weight $2$. One can construct now supersymmetry invariant terms in the action as  chiral integrals $\int d^{4}\theta$ with arbitrary polynomials involving the  kinetic multiplet and other chiral multiplets. Since  four antichiral derivatives have the same effect as the four antichiral integrals,  these terms  correspond to  D-terms with non chiral integrals $\int d^{4 }\theta d^{4} \bar \theta$  of terms involving the original field $\Phi$. The nonrenormalization theorem of  \cite{deWit:2010za}  shows that D-terms of this type do not contribute to the Wald entropy.
Since the renormalized action of the localizing instantons follows from the bulk action and has the same form as the entropy function, it should be possible to extend this nonrenormalization theorem to the renormalized action discussed in this paper. 
\end{myitemize}
These two points indicate that the D-terms, or at least a large subclass of them,  may in fact not contribute to the renormalized action. 

Adding hyper multiplets does not change the transformation rules of the vector multiplets. We therefore expect that  the localizing instantons that we have found here will continue to exist. There could be in principle additional localizing solutions where hyper multiplet fields are excited but this may not necessarily happen. It then only remains to check that the coupling of hyper multiplets and vector multiplets at high order cannot contribute to the renormalized action. Lacking an offshell formulation of couplings between hypers and vectors, we cannot at present address this question but perhaps something analogous to the nonrenormalization theorem discussed above can be extended to these terms as well.

In any case, these questions  can be systematically investigated in the context of our off-shell localizing instantons. If some of the D-terms do happen to contribute to the renormalized action, their contribution can  be taken into account by evaluating them on  the off-shell solutions. Similarly if there are new localizing instantons upon the inclusion of hypers, those too can be added as separate contribution to the final answer for  the functional integral. 

We would like to add that in several cases such as small black holes and  big black holes in models with $\CN =4$ supersymmetry, exact microscopic degeneracies are known \cite{Dijkgraaf:1996it, Gaiotto:2005gf, David:2006yn, Dabholkar:2008zy, Dabholkar:1989jt, Dabholkar:1990yf}.  One can thus apply the formalism developed here in these specific cases to test   the macroscopic results against a known answer from the microscopic side. These applications of our results will be reported in a forthcoming publication\cite{Dabholkar:2010t}. We will only note here that in all these models,  the macroscopic answer obtained ignoring D-terms and hyper multiplets appears to agree in remarkable details with the exact microscopic answer.  The agreement between microscopic and macroscopic answers thus gives additional evidence that  ignoring D-terms  and hypermultiplets may be justified at least with $\CN=4$ supersymmetry. Unfortunately, a manifestly off-shell formalism is not available for the $\CN=4$ theory. So at present,
 it remains  an interesting open problem in offshell supergravity to check to what extent the  assumption of ignoring D-terms and hypermultiplets is  justified.

If the hyper multiplets and D-terms can be ignored for reasons outlined above, one can conclude that  
$W_{0}(q, p)$ has the same form as $\hat W(q, p)$ evaluated  in \S\ref{Solution} 
 \begin{equation}\label{integral3}
 W_{0} (q, p) = \int_{\mathcal{M}_{Q}} \, e^{- \pi  \phi^{I} q_{I}}  \, | Z_{top}(\phi, p ) |^{2}  \, 
 \, Z_{det}  \, [dC]_{\mu} \end{equation}
The contribution from the orbifolds of $AdS_{2}$ also has a very similar structure since the localizing instanton solution is still valid.

\subsection{Non-perturbative corrections from orbifolds of $AdS_{2}$ \label{nonpert}}

As we have seen the quantum entropy receives contributions  not only from local higher-derivative terms in the effective 
action via the Wald formula but also from nonlocal effects arising from integrating out massless fields.  As discussed,
upon localization,  these nonlocal effects are entirely captured by the ordinary integrals over 
the parameters $\{C^{I}\}$ and $Z_{det}$. The nonperturbative corrections due to brane instantons are  included in the integrand by $|Z_{inst}|^{2}$.

This is, however,  not the whole story. There are additional nonperturbative corrections which arise from  subleading gravitational saddle points with the 
same $AdS_{2}$ boundary conditions but whose action is exponentially suppressed given by the terms with positive $s$ in \eqref{Wsum}.   Localization allows us to 
evaluate the contribution for each family of these subleading saddle points as well, with a structure similar to the 
leading saddle points. Even though these saddle points are exponentially suppressed, it is meaningful to keep them 
relative to the power-law suppressed terms, because the evaluation of the functional integral using localization is 
exact and not an asymptotic expansion.


Based on the structure of known microscopic degeneracy formulas, 
 it was proposed in \cite{Banerjee:2008ky,Sen:2009vz} that there is a universal series of 
saddle points to the path integral semiclassical solutions 
of the theory with the same boundary conditions as the pure $AdS_{2}$, but which differs in the bulk. 
These saddle points give  non-perturbative 
corrections to the entropy of the form $e^{\CE_{0}/c}$ for $c=2,3\dots$ 
These solutions are orbifolds of $AdS_{2}\times M$, where the 
orbifold action on the $AdS_{2}$ factor of the geometry is   
a $\IZ_{c}$ quotient in the angular direction of Euclidean $AdS_{2}$:
\be\label{orbads2}
d{s'}^2   =  v  \left[ ({r'}^2-1) \, d{\theta'}^2 + \frac{d{r'}^2}{{r'}^2-1} \right] , \qquad \theta' \sim \theta' + 2 \pi  /c \ . 
\ee
so that indeed the bulk action gets reduced by a factor of $c$ compared to the original solution $c=1$. 
By a change of coordinates $r'=c r, \theta'=\theta/c$, one gets 
\be\label{orbads2new}
d {s}^{2} = v \left[ 
\left({r}^2-\frac{1}{c^2}\right ) d \theta^2+ \frac{d{r}^2}{{r}^2-\frac{1}{c^2}}  \right]\ ,\qquad \theta \sim \theta + 2 \pi \ ,
\ee
which makes it manifest that the asymptotics of all the solutions are the same as 
required \eqref{asympcond}.

These solutions of course have a conical singularity at the center of $AdS_{2}$, which we do not 
{\it a priori} expect to be included in the sum over configurations. To remedy this, \cite{Banerjee:2008ky} 
proposed that only such singularities which can be resolved in string theory should be allowed in this sum.
This includes orbifolds which, in addition to the above action, simultaneously act as a rotation on the 
internal $S^2$ which is generically present in four dimensional black holes. In this case, the singularity 
is locally of the $\IC^{2}/Z_{n}$ type which is known to be resolvable in string theory after the addition of a 
B field on a collapsing cycle as in the case of usual orbifold singularities.

In \cite{Murthy:2009dq}, it was noticed that there is another series of saddle points associated with 
each electric charge that the black hole carries\footnote{The second type of orbifold can be called `electric' in that it can be understood 
simply as an insertion of a Wilson line associated with an electric charge, along with 
a geometric action on the $AdS_{2}$. The first type can be called `magnetic'  in that it can be understood in similar terms as the insertion 
of an 't Hooft line corresponding to a magnetic charge \cite{Murthy:2009dq}.}. 
The construction is as follows -- corresponding to each electric charge $q_{i}$, one has a 
gauge potential $A_{i}$ which classically is of the form \eqref{ggefld}. Along with the above action 
\eqref{orbads2}, if one simultaneously makes a gauge transformation with a constant parameter
(thus generating  a constant gauge field at asymptotic infinity), then the orbifold action has no fixed 
point in the full configuration space which includes the geometry and all the Wilson lines, and therefore 
there is no singularity. In the Euclidean theory, if the orbifold action respects the supersymmetries, 
then it must also involve a rotation in the $S^{2}$~\cite{Banerjee:2008ky}. 

%


These smooth orbifolds are labelled (for each charge $q_{i}$) by a pair of integers $(c,d_{i})$ and $c \ge 1$. 
Changing $d\to d+c$ does not change the Wilson line, and so one has $1 \le d_{i} < c$. In addition, one must 
demand that there are no new fixed points that arise, and this gives the condition that $(c,\{ d_{i} \})$ 
are relatively prime. 

In the case that the electric charge $q_{i}$ has an interpretation as a momentum on a compact circle in the 
internal geometry $M$, then the above construction gets a geometric life in the three dimensional theory 
as the very near-horizon limit of a family of extremal BTZ black holes with the same mass and charge 
\cite{Murthy:2009dq}. 
In this case, we can think of the inequivalent solutions are
labelled by double cosets $\Gamma_\infty \backslash SL(2,\IZ) / \Gamma_\infty$
where 
$$\Gamma_\infty = {\begin{pmatrix} 1 & * \\ 0 & 1 \end{pmatrix}}\, .$$
This family can be thought of as related to a different family of black holes in asymptotically $AdS_{3}$ 
space with fixed electric potential \cite{Maldacena:1998bw} by a Laplace transform in spacetime. 
These two families of orbifolds together were sufficient to give a semiclassical interpretation of the 
microscopic partition function of supersymmetric black holes in $\CN=4$ theories. 

In the $\CN=2$ situation, there are no geometric one-cycles in $M$ in the weak coupling limit (although there is 
always one $S^{1}$ in the M-theory frame). However, the general construction sketched above applies for 
each of the electric charges which arise from branes wrapping cycles inside the Calabi-Yau, and each such
charge should give rise to a family of saddle points. In general,  these families may not all be independent, 
and may be related by the duality symmetries that preserve a given charge configuration. 
This is indeed the case in the $\CN=4$ theory where we know the microscopic answer --
there is only one family of solutions for each T-duality invariant. 
From the $\CN=4$ answer, it seems like the 't Hooft line construction 
applies to each of the magnetic charges as well (up to T-duality as above), but S-duality does not correlate the  
electric and magnetic family of solutions, and they have independent labels.

\subsection{Relation to classical entropy function and the topological string \label{Relation}}

We would like to conclude with a few comments about the relation of our results with the topological string and earlier works.

\begin{myitemize}
\item One of the remarkable facts about the formula \eqref{integral3} is that the renormalized action that appears in the integrand is precisely the classical
entropy function  as $\exp [\mathcal{E}(\phi, p)]$ \cite{Sen:2007qy}. As mentioned in the introduction,  the classical entropy function $\CE$ is simply an elegant way 
to summarize the classical attractor equations. It  is simply a  Legendre transform of the classical action
with respect to the electric charges, and its value at the critical  point equals the Wald entropy. 
Since the definition of the entropy function is purely classical, only the critical points of this function and the 
value of the function at the critical point have any physical meaning. 
This fixes only two terms in the Taylor expansion around the critical point and thus there are  any number of functions with the same critical point and the same value at the critical point. It is not clear which will play the role of an off-shell action.  Ooguri, Strominger, and Vafa  of 
\cite{Ooguri:2004zv} made an inspired guess to attribute meaning to the entropy  function away from critical point and 
elevated it  to an off-shell action of  a mixed ensemble. This means that to obtain the degeneracy, one must  
effectively elevate the classical Legendre transform to an inverse Laplace transform. 
Our derivation of the localization action explains from first principles why such an off-shell action can make sense. 
Moreover, it gives a systematic way to determine contributions from one-loop determinants and brane-instantons. The integration measure is inherited from the supergravity measure and since the range of $C$ fields is the entire real line  the $\phi$ contours of integral are parallel to imaginary axis as for any inverse Laplace transform.

\item 
If we ignore $|Z_{inst}|^{2}$ and $Z_{det}$ in \eqref{integral3} then   $|Z_{top}(\phi, p ) |^{2}$ could be regarded precisely as the  mixed ensemble partition function of OSV conjecture with the measure for the inverse Laplace transform understood to be determined as above. A  derivation of this conjecture has been outlined in  \cite{Gaiotto:2006ns} which uses $AdS_{3}$ geometry and dilute gas approximation. This is justified in the region where the MSW picture gives the complete result and for range of charges for which dilute gas of M2-branes and anti M2-branes dominates the path integral. By contrast, methods outlined in this paper provide a way to derive the quantum entropy for arbitrary charges without any restriction.

\item A derivation of the $|Z_{top}(\phi, p ) |^{2}$ factor  for all orders in perturbation theory in a 
large charge expansion  has been suggested 
in \cite{Beasley:2006us, Chandrasekhar:2008qx} using world sheet string theory in the $AdS_{2}$ background. In this picture 
worldsheet instantons at the north pole contribute $Z_{top}$ whereas worldsheet instantons at the 
south pole contribute $\bar Z_{top}$ giving a new perspective on the form of the integrand. These 
string worldsheet computations are necessarily tied to a fixed classical background which solves the 
string equations of motion. 
As  we have seen, the $AdS_{2}$ classical background does not access arbitrary values of 
the electric potential $\phi$ which are fixed to the attractor value. Our localizing instanton solution provides a way access the large field regions of the functional integral away
from the critical point. However  perturbative worldsheet calculations would not be applicable around  such off-shell configurations and it is necessary to evaluate the supergravity functional integral.

\item The $AdS_{2}$ boundary conditions require to hold fixed the  charges and the electric fields at the boundary have to held fixed at their attracror value.  Moreover, the integration variables 
are  not really the electric fields $e^{I}$ conjugate to the charge $q_{I}$ but rather the parameters 
$C^{I}$ which is the value of  the auxiliary fields at the origin of the $AdS_{2}$. Indeed, integration over the electric field as suggested by the OSV conjecture  presented an important conceptual difficulty since it would appear to be in 
conflict with the usual rules of  $AdS_{2}$ holography.  Such an integration would imply integrating over different boundary conditions. Our derivation of the 
integral \eqref{integral2} shows that the electric field indeed remains fixed at the attractor values.  The integration 
over the parameters $C$ enters into the story for entirely different reasons having to do with localization.  
All auxiliary fields for the localizing solutions vanish at the boundary and thus respect the $AdS_{2}$ boundary 
conditions.

\item As we have seen in \S\ref{nonpert}, in string theory there are additional nonperturbative corrections that arise from gravitational saddle 
points with the same $AdS_{2}$ boundary conditions. These can be viewed as orbifolds of the leading solution.
As a result, their contribution has a very similar structure to \eqref{integral3}. 
These contributions also indicate $W(q, p)$ in a microcanonical ensemble is a more natural object than the  integrand of \eqref{integral2} which could be thought of the partition function in a mixed ensemble. Conversely, one can use the knowledge of the microscopic answer to figure out the rules of nonperturbative quantum gravity to determine which configurations have to be included in the functional integral. 

\item The quantum degeneracies of black holes $d(q, p)$ in an appropriate duality frame are given by the Donaldson-Thomas invariants. Given an exact evaluation of the same quantity from macroscopic side in terms of $W(q, p)$  involves the topological string partition function which is related to the Gromov-Witten invariants. Results in this paper can provide a way to establish a precise relation between these two very different counting problems. Note however that $W(q, p)$ gives the degeneracies of a single black hole horizon. To compare with the microscopic side it is necessary to separate the contribution from single-centered black holes to extract from the Donaldson-Thomas invariants which count the quantum states in asymptotically flat spacetime. This is an interesting mathematical problem in itself. In an analogous $\CN=4$ situation a complete solution of this problem is known \cite{Dabholkar:2012nd}.  It may be possible in simple examples to arrive at  `exact holography' where both bulk and boundary 
partition functions can be computed exactly \cite{Dabholkar:2010t}. This connection could be used conversely  to figure out the rules of nonperturbative functional integral of quantum gravity and string field theory in $AdS_{2}$ using the knowledge of the black hole degeneracies since  one then knows what the functional integral must evaluate to.  Perhaps there is a `twisted' version of this  functional integral of string field theory that will focus directly on this `topological' BPS sector of the theory.

\end{myitemize}

\subsection*{Acknowledgments}

We would like to thank Chris Beasley, Bernard de Wit, Nadav Drukker, Vasily Pestun, and Amir Keshani-Poor for useful discussions. We are especially grateful to Ashoke Sen for several illuminating discussions throughout the course of this work. The work of A.~D. was
supported in part by the Excellence Chair of the Agence Nationale de la
Recherche (ANR).
The work of J.~G. is supported in part by Fundac\~{a}o para Ci\^{e}ncia e
Tecnologia (FCT). The work of S.~M. is supported in part by the European 
Commission Marie Curie Fellowship under the contract PIIF-GA-2008-220899. 

\appendix

\section{Killing spinors in the attractor geometry \label{Killing}}

To apply localization arguments, it is necessary to identify the supercharge $Q$ that squares to the compact bosonic generator $L - J$. For this purpose, it is useful to know first the explicit form of the Killing spinors in the on-shell attractor geometry. 

Recall that in the superconformal formalism, there are fermionic variations corresponding to $Q$ as well as $S$, 
which we label by $\ve$ and $\eta$ respectively \cite{Mohaupt:2000mj}. One can only impose $Q$-invariance up to a uniform 
$S$-supertranslation. This corresponds to the fact that the physical supersymmetries in the Poincar\'e theory are 
found after the gauge fixing procedure to be a linear combination of these two variations. 
In general, this combination has a complicated dependence on the other fields as well as the choice of prepotential. 
The method of \cite{LopesCardoso:2000qm} is to surpass this problem by finding spinor fields whose variation under $S$ vanishes. 
One can then simply use the $Q$-invariance conditions for these spinor fields, which by construction is gauge 
independent. This construction was very useful in \cite{LopesCardoso:2000qm} to find the half-BPS solution in asymptotically flat space.

However, these gauge-independent supersymmetry transformations then depend on the choice of prepotential and hence the choice of the Lagrangian. This  is not well-suited for our purposes since we are really interested in the  off-shell localizing solutions that are determined direcly by the off-shell supersymmetry transformation without any reference to the prepotential. Moreover, we are only interested in the near horizon geometry which is much simpler to analyze than the full black hole solution including the asymptotic infinity. For the near horizon supersymmetries,  we make the simple observation that a choice of the bosonic fields corresponding to the near horizon attractor 
background leads to a particularly simple choice of gauge for the physical theory, namely $\eta=0$. 
This choice then permits us to work with the simpler supersymmetry transformations of the superconformal theory. 

To see this, we begin by imposing the vanishing of the variations of fermionic fields of the Weyl mutiplet:
\begin{eqnarray} \label{Weylvar1}
   0 =   \delta \psi^i_{\mu} & = & 2D_{\mu} \epsilon^i-\frac{1}{8}\gamma_a\gamma_b T^{abij}\gamma_{\mu} \epsilon_j+
      \gamma_{\mu}\eta^i \ , \\
 \label{Weylvar2} 
    0 =  \delta\chi^i & = & -\frac{1}{12}\gamma_a\gamma_b\displaystyle{\not}DT^{abij} \epsilon_j+D \epsilon^i
      +\frac{1}{12}T^{ij}_{ab}\gamma^a\gamma^b\eta_j \ , \\
  \label{Weylvar3}   
   0 =  \delta\phi^i_{\mu} & = & -2f^a_{\mu}\gamma_a \epsilon^i-\frac{1}{4}\displaystyle{\not}DT^{ij}_{cd}\sigma^{cd}
      +2D_{\mu}\eta^i \ .
\end{eqnarray}
At the attractor values, we have 
\be \label{attval}
v=\frac{16}{\omega\bar{\omega}} \ , \qquad  T_{rt}^-= v \omega \ ,
\ee
and the above variations simplify to 
\begin{eqnarray} \label{Weylvar22}
 \delta \psi^i_{\mu} & =& 2D_{\mu} \epsilon^i-\frac{1}{8}\gamma_a\gamma_b T^{abij}\gamma_{\mu} \epsilon_j+\gamma_{\mu}\eta^i \ , \\
      \delta\chi^i & =& \frac{1}{12}T^{ij}_{ab}\gamma^a\gamma^b\eta_j \ , \\
      \delta\phi^i_{\mu} & =& 2D_{\mu}\eta^i \ .
\end{eqnarray}

From here, we deduce the $AdS_2\times S^2$ Killing spinors equations
\begin{eqnarray} \label{killing eq}
 D_{\mu} \epsilon^i & =& \frac{1}{16}\gamma_a\gamma_b T^{abij}\gamma_{\mu} \epsilon_j  \ , \nonumber\\
  D_{\mu} \epsilon_i & =& \frac{1}{16}\gamma_a\gamma_b T^{ab}_{\quad ij}\gamma_{\mu} \epsilon^j\\
 \eta_i=\eta^{i} & = & 0 \nonumber . 
\end{eqnarray}
We thus see that $\eta^{i}=0$ as promised. Before solving the equation for $ \epsilon^{i},\epsilon_i$, 
note that in the Euclidean theory in four dimensions, the spinors should have a symplectic-Majorana condition imposed on them, while in Minkowski spacetime they can be majorana or symplectic-Majorana \cite{Cortes:2003zd}. In addition, the Weyl projection is not compatible with the majorana condition in the Minkowski case and therefore the left and right-handed spinors are complex conjugate to each other. On the contrary, in the Euclidean case, we can have symplectic Majorana-Weyl spinors but not majorana
\begin{equation} \label{Majsymp}
(\zeta^{i}_{\pm})^{*}=-i \ve_{ij} \, (\sigma_1\times \sigma_2) \,\zeta^{j}_{\pm},
\end{equation}
where the indices $i,j$ are $SU(2)'$ quantum numbers and $\ve_{ij}$ is the antisymmetric tensor of $SU(2)'$. In the literature \cite{Mohaupt:2000mj} the spinors used obeyed a majorana condition in Minkowski space. They used the convention that positive/negative chirality is correlated with upper/down $SU(2)'$ indice due to complex conjugation. Since the killing spinor equations \ref{killing eq} were derived from the Lorentzian theory, we shall use an ansatz which reproduces the killing spinor equations in Euclidean $AdS_2\times S^2$. The ansatz is the following
\begin{eqnarray}\label{ansatz}
 \epsilon_i=i\ve_{ij}\xi^j_{-}\\
  \epsilon^i=\xi^i_{+}\\ 
\nonumber
\end{eqnarray}Note that we explicitly show the chirality of the spinor. We should therefore solve the Killing spinor condition for an unconstrained Dirac spinor $\xi^i=\xi^i_{+}+\xi^i_{-}$, double the space and 
then impose the above constraint \eqref{Majsymp}. 
We represent the Dirac spinor $ \xi$ as a direct product $ \xi= \xi_{AdS_2}\otimes \xi_{S^2}$ 
where $ \xi_{AdS_{2}}$ and $ \xi_{S^{2}}$ are two component spinors,  
and use the following gamma matrix representation 
\begin{equation}
\gamma_{\theta}= \sqrt{v} \, \sinh\eta\,\sigma_1\otimes 1\ , \quad \gamma_{\eta}=  \sqrt{v} \, \sigma_2\otimes
1\ , \quad \gamma_{\phi}=  \sqrt{v} \, \sin\psi\, \sigma_3\otimes \sigma_1\ , \quad 
\gamma_{\psi}=  \sqrt{v} \, \sigma_3\otimes \sigma_2 \ ,
\end{equation}
where $v \equiv v_{1} (=v_{2})$ is the classical size of the $AdS_{2}$ (and the $S^{2}$). 

Equations \eqref{killing eq} simplify to the diagonal form 
\begin{eqnarray}
 D_{\mu} \xi^i_{AdS_2} & = & \frac{\omega}{|\omega|}\frac{i}{2}(\sigma_3\times 1) \, \gamma_{\mu} \,  \xi^i_{AdS_2} \ , \\
 D_{j} \xi^i_{S^2} & = & \frac{\omega}{|\omega|}\frac{i}{2}(\sigma_3\times 1) \, \gamma_{j} \,  \xi^i_{S^2} \ . \\
\end{eqnarray}
which are easily solved \cite{Lu:1998nu}. In the bispinor basis
\begin{eqnarray}
  \xi&=& a_1\left(\begin{array}{c}
            1\\ 
	    0
           \end{array}\right)\times\left(\begin{array}{c}
            1\\ 
	    0
           \end{array}\right)+a_2\left(\begin{array}{c}
            0\\ 
	    1
           \end{array}\right)\times\left(\begin{array}{c}
            1\\ 
	    0
           \end{array}\right)
	    +a_3\left(\begin{array}{c}
            1\\ 
	    0
           \end{array}\right)\times\left(\begin{array}{c}
            0\\ 
	    1
           \end{array}\right)+a_4\left(\begin{array}{c}
            0\\ 
	    1
           \end{array}\right)\times\left(\begin{array}{c}
            0\\ 
	    1
           \end{array}\right)\nonumber\\ \nonumber\\
&\equiv&\left(\begin{array}{c}
            a_1\\
	    a_2\\	
	    a_3\\	
	    a_4
          \end{array}\right)\nonumber \\
\end{eqnarray}the solutions are (this is assuming that $w \in \IR^{+}$, and we have fixed a certain normalization for the spinors):
\begin{eqnarray} \label{Killspin1}
 \xi^i_{--}= 2 \, e^{-\frac{i}{2}(\theta+\phi)}\left(\begin{array}{c}
\cosh\frac{\eta}{2}\cos\frac{\psi}{2}\\
                                                         \sinh\frac{\eta}{2}
\cos\frac{\psi}{2}\\
                                                        
-\cosh\frac{\eta}{2}\sin\frac{\psi}{2}\\
                                                        
-\sinh\frac{\eta}{2}\sin\frac{\psi}{2}\end{array}\right)  & \ , \qquad & 
 \xi^i_{-+} = 2 \, e^{-\frac{i}{2}(\theta-\phi)}\left(\begin{array}{c}
\cosh\frac{\eta}{2}\sin\frac{\psi}{2}\\
                                                         \sinh\frac{\eta}{2}
\sin\frac{\psi}{2}\\
                                                        
\cosh\frac{\eta}{2}\cos\frac{\psi}{2}\\
                                                        
\sinh\frac{\eta}{2}\cos\frac{\psi}{2}\end{array}\right) \ , \nonumber \\
 \xi^i_{+-} = 2 \, e^{\frac{i}{2}(\theta-\phi)}\left(\begin{array}{c}
\sinh\frac{\eta}{2}\cos\frac{\psi}{2}\\
                                                         \cosh\frac{\eta}{2}
\cos\frac{\psi}{2}\\
                                                        
-\sinh\frac{\eta}{2}\sin\frac{\psi}{2}\\
                                                        
-\cosh\frac{\eta}{2}\sin\frac{\psi}{2}\end{array}\right) & \ , \qquad &  
 \xi^i_{++} = 2 \, e^{\frac{i}{2}(\theta+\phi)}\left(\begin{array}{c}
\sinh\frac{\eta}{2}\sin\frac{\psi}{2}\\
                                                         \cosh\frac{\eta}{2}
\sin\frac{\psi}{2}\\
                                                        
\sinh\frac{\eta}{2}\cos\frac{\psi}{2}\\
                                                        
\cosh\frac{\eta}{2}\cos\frac{\psi}{2}\end{array}\right) \nonumber \\ . 
\end{eqnarray}As explained above, we should impose a symplectic-Majorana conditon on the spinors.
In the above basis, equation \eqref{Majsymp} implies: 
\begin{eqnarray}
 &&\xi^{+}_{++}=(\xi^{-}_{--})^* \nonumber\\
 &&\xi^{+}_{-+}=(\xi^{-}_{+-})^*\nonumber\\
 &&\xi^{-}_{++}=(-\xi^{+}_{--})^*\nonumber\\
 &&\xi^{-}_{-+}=(-\xi^{+}_{+-})^*\nonumber
\end{eqnarray}
where the star is not the ordinary complex conjugation but the complex
conjugation condition as defined by the symplectic-majorana condition.

One can now identify the spinors $ \epsilon^{i}_{ra}$ as the generators of  $G^{ia}_{r}$, the supercharges 
of the near horizon $\CN=4$ superalgebra \S\ref{Supersymmetries}. 
The real combinations $Q_{\mu}, \wt Q_{\mu}, \m = 1,\dots , 4$ are generated by the combinations:
\begin{eqnarray} \label{defzeta}
\begin{array}{l}
\zeta_1= \xi^+_{++}+ \xi^-_{--},\\
\zeta_2=-i\left( \xi^+_{++}- \xi^-_{--}\right),\\
\zeta_3=-i\left( \xi^-_{++}+ \xi^+_{--}\right),\\
\zeta_4= \xi^-_{++}- \xi^+_{--},\end{array}
\,\,
\begin{array}{l}
\tilde{\zeta}_1= \xi^+_{-+}+ \xi^-_{+-},\\
\tilde{\zeta}_2=-i\left( \xi^+_{-+}- \xi^-_{+-}\right),\\
\tilde{\zeta}_3=-i\left( \xi^-_{-+}+ \xi^+_{+-}\right),\\
\tilde{\zeta}_4= \xi^-_{-+}- \xi^+_{+-},\end{array}
\end{eqnarray}We can easily see that these killing spinors are real under the
complex conjugation condition defined by \eqref{Majsymp}. As an instructive
exercise take for example $\zeta^1$. The $SU(2)'$ components are
$\zeta^{1+}=\xi^{+}_{++}$ and $\zeta^{1-}=\xi^{-}_{--}$. Both are
complex conjugate to each other
\begin{eqnarray}
 (\zeta^{1+})^*=-i \ve_{+-} \, (\sigma_1\times \sigma_2) \, \zeta^{1-}
\nonumber\\
  (\zeta^{1-})^*=-i \ve_{-+} \, (\sigma_1\times \sigma_2) \, \zeta^{1+}\nonumber
\end{eqnarray}

Recall that the supersymmetry variations for fermions and scalars of the vector multiplets in Minkowski theory are \cite{Mohaupt:2000mj}
\begin{eqnarray}
 &&\delta X^I=\bar{\epsilon}^i\Omega^I_i \nonumber\\
  &&\delta \bar{X}^I=\bar{\epsilon}_i\Omega^{Ii} \nonumber\\
&& \delta \Omega^I_i=2\displaystyle{\not} \partial X^I\epsilon_i+\frac{1}{2}\ve_{ij}\mathcal{F}^{I\mu\nu-}\gamma_{\mu}\gamma_{\nu}\epsilon^j+Y^{I}_{ij}\epsilon^j+2X^I\eta_i \nonumber\\
&& \delta \Omega^{Ii}=2\displaystyle{\not} \partial \bar{X}^I\epsilon^i+\frac{1}{2}\ve^{ij}\mathcal{F}^{I\mu\nu+}\gamma_{\mu}\gamma_{\nu}\epsilon_j+Y^{Iij}\epsilon_j+2\bar{X}^I\eta^i\nonumber
\end{eqnarray} where  $\Omega_i$ has positive chirality while $\Omega^i$ has negative chirality. Changing basis from the $\epsilon$ spinors to the $\zeta$ spinors using (\ref{ansatz}),we can reexpress the susy variations  as
\begin{eqnarray}\label{susy variantions}
 &&\delta X^I=-(\zeta^i_{+})^{\dagger}\lambda^{Ii}_{+} \nonumber\\
  &&\delta \bar{X}^I= -(\zeta^i_{-})^{\dagger}\lambda^{Ii}_{-}\nonumber\\
&& \delta \lambda^{Ii}_{+}=\frac{1}{2}(F_{\mu\nu}^{I-}-\frac{1}{4}\bar{X}^{I} \, T^{-}_{\mu\nu}) \, 
 \gamma^{\mu} \, \gamma^{\nu} \, \zeta^{i}_+ +2i \displaystyle{\not}\partial X^{I} \, \zeta^i_-+Y^{Ii}_j \, \zeta^j_+ \nonumber\\
&& \delta \lambda^{Ii}_{-}=\frac{1}{2}(F_{\mu\nu}^{I+}-\frac{1}{4}X^{I} \, T^{+}_{\mu\nu}) \, 
 \gamma^{\mu} \, \gamma^{\nu} \, \zeta^{i}_- +2 i \displaystyle{\not}\partial \bar{X}^{I} \, \zeta^i_+ +Y^{Ii}_j \, \zeta^j_- \,
\end{eqnarray}
where $\lambda$ are related to $\Omega$ spinors by 
\begin{equation}
\Omega_{i } = \varepsilon_{ij} \lambda^{j}_{-} \qquad \Omega^{i} = -i \lambda^{i}_{+} \, .
\end{equation}

Under a transformation generated by $\zeta_i$ or $\tilde{\zeta}_i$, given in (\ref{defzeta}), we can show that the action of $\delta^2$ is $L-J$ or $L+J$ respectively
\begin{eqnarray}
 \delta^2X^I=-(\zeta^i_{+})^{\dagger}\delta \lambda^{Ii}_{+}=2i(\zeta^i_{+})^{\dagger} \displaystyle{\not}\partial X^{I} \, \zeta^i_-\\
\delta^2\bar{X}^I=-(\zeta^i_{-})^{\dagger}\delta\lambda^{Ii}_{-}=2i(\zeta^i_{-})^{\dagger}\displaystyle{\not}\partial \bar{X}^{I} \, \zeta^i_+
\end{eqnarray}where the remaining contractions vanish identically for the spinors chosen. After a straightforward computation we find 
 \begin{eqnarray}
 \delta^2X^I=-2i (\partial_{\theta}-\partial_{\phi}) X^{I} =2(L-J)X^I\\
\delta^2\bar{X}^I=-2i (\partial_{\theta}-\partial_{\phi}) \bar{X}^{I}=2(L- J)\bar{X}^I \, .
\end{eqnarray}

\section{Localization equations}\label{appB}
The system of coupled first order differential equations that we want to solve is
\begin{eqnarray}
 && \cosh(\eta)H+\cos(\psi)J+\partial_1 H\sinh(\eta)+\partial_3 J\sin(\psi)-\frac{K}{2}=0 \label{eq1app}\\
&& \cosh(\eta)J+\cos(\psi)H-\partial_1 J\sinh(\eta)-\partial_3 H\sin(\psi)-\frac{K}{2}\cos(\psi)\cosh(\eta)=0\label{eq2app}\\
&& \partial_1 H \sin(\psi)=\partial_3 J \sinh(\eta)-\frac{K}{2}\sin(\psi)\sinh(\eta)\label{eq3app}\\
&& \partial_3 H \sinh(\eta)=\partial_1 J \sin(\psi)\label{eq4app}
\end{eqnarray}
together with $\partial_{0,2}H=\partial_{0,2}J=0$.
Using equations (\ref{eq3app}) and (\ref{eq4app}) in (\ref{eq1app}) and (\ref{eq2app}), we obtain
\begin{eqnarray}
 &&\cosh(\eta)H+\cos(\psi)J+\partial_1 H\frac{\cosh(\eta)^2-\cos(\psi)^2}{\sinh(\eta)}-\frac{K}{2}\cos(\psi)^2=0 \label{eq26}\\
&& \cosh(\eta)J+\cos(\psi)H-\partial_1 J\frac{\cosh(\eta)^2-\cos(\psi)^2}{\sinh(\eta)}-\frac{K}{2}\cos(\psi)\cosh(\eta)=0
\end{eqnarray}Multiplying the first equation by $\cosh(\eta)$, the second by $\cos(\psi)$ and subtracting after, we find
\begin{equation}
 \partial_1(H\cosh(\eta)+J\cos(\psi))=0
\end{equation}This equation gives the integrability condition $H\cosh(\eta)+J\cos(\psi)=f(\psi)$, where $f(\psi)$ is some undetermined function. Repeating the same exercise but now for $\partial_3H$ and $\partial_3 J$ we find
\begin{equation}
 \partial_3(H\cosh(\eta)+J\cos(\psi))=0
\end{equation}which then implies $\partial_{\psi}f(\psi)=0$. In other words, $f(\psi)$ is a constant that we label as $C$.

Before proceeding with the general solution let's solve the equations for the case $K=0$. In other examples of localization we saw that new solutions can arise when the auxiliary fields are excited. Therefore we expect to find the vacuum solution $H=J=0$ for $K=0$. In the following we show that this is indeed the case.

From equation (\ref{eq26}) we have:
\begin{equation}
 \partial_1 H=-C\frac{\sinh(\eta)}{\cosh(\eta)^2-\cos(\psi)^2}
\end{equation}with $H\cosh(\eta)+J\cos(\psi)=C$. We easily see that $\partial_1 H$ is singular at the points $\eta=0,\psi=(0,\pi)$. Since we are looking for smooth fluctuations which preserve the boundary conditions, we are forced to set $C=0$ and therefore $H=0$ and $J=0$. This can be easily seen from the fact that for $C=0$ $\partial_1H=0$ implies $H=g(\psi)$, where $g(\psi)$ is some function. This solution does not respect the boundary conditions unless $g(\psi)=0$.

In the case of non-zero $K$ we can have additional solutions. In fact the solution 
\begin{equation}
 H=\frac{C}{\cosh(\eta)},\;\;J=0,\;\;K=\frac{2C}{\cosh(\eta)^2}
\end{equation}nicely solves all equations. Let's see that this is the case. 

We start by assuming an ansatz solution for $H$ and $J$ of the form
\begin{equation}
 H=\frac{C}{\cosh(\eta)}+\sum_{k\geq 2} \frac{a_k(\psi)}{\cosh(\eta)^k},\;\; J=\sum_{k\geq 1}\frac{b_k(\psi)}{\cosh(\eta)^k}
\end{equation} where $a_k(\psi)$ and $b_k(\psi)$ are smooth arbitrary functions. The constraint $H\cosh(\eta)+J\cos(\psi)=C$ implies the following recursive relation
\begin{equation}
 a_{k+1}(\psi)+b_k(\psi)\cos(\psi)=0,\; k\geq 1.
\end{equation}This together with equation (\ref{eq4app}) gives the following differential equation for $a_k(\psi)$
\begin{equation}
 \partial_{\psi}a_k(\psi)=k\frac{\sin(\psi)}{\cos(\psi)}a_k(\psi),\; k\geq 2.
\end{equation}Solving this differential equation is a straightforward exercise, giving
\begin{equation}
 a_k(\psi)=\frac{\tilde{C}}{|\cos(\psi)|^k}
\end{equation}which is singular along the equator of the sphere , that is, for $\psi=\pi/2$. We are then forced to set $\tilde{C}=0$ which gives $a_k=0$ for all $k$.

We finally conclude that the only smooth solution which respects the boundary conditions is then,
\begin{equation}
 H=\frac{C}{\cosh(\eta)},\;\;K=\frac{2C}{\cosh(\eta)^2},\;\; J=0
\end{equation}

In the meanwhile we came to know that the authors of \cite{Gupta:2012cy} also found the same solutions using a different method.

\section{Some aspects of the superconformal multiplet calculus \label{susyvar}}

In this appendix, we shall summarize some aspects of the superconformal multiplet calculus 
which we briefly presented in \S\ref{off-shell}. We shall first present the supersymmetry 
variation of the various multiplets. We shall then present the invariant Lagrangian density formula 
for a chiral multiplet. We shall then present the rule which defines the various
components of a scalar function of chiral superfields {\it e.g.} the prepotential superfield 
$\bf F (\bf X^{I})$. These are the basic ingredients that go into building the superconformal action. 
We shall borrow the presentation of the recent \cite{deWit:2010za} wherein a lot of these 
facts (and more) have been collected, this can be referred to for more details. 

The invariance of the bulk Lagrangian under the superconformal transformations are 
well established, we provide these details for the sake of completeness. Using the same transformations, 
in another appendix, we shall sketch the supersymmetry invariance of 
our conjectured boundary action. This, as far as we know, is new, and there is scope to develop it further. 

As in the text, $\epsilon_{i}$ and $\eta_{i}$ denote the parameters of the $Q$ and $S$ supersymmetry 
transformations. The transformation rules for a chiral multiplet of Weyl weight $w$ are:
\begin{align}
  \label{eq:conformal-chiral}
  \delta A =&\,\bar\epsilon^i\Psi_i\,, \nonumber\\[.2ex]
  \delta \Psi_i =&\,2\,\Slash{D} A\epsilon_i + B_{ij}\,\epsilon^j +
  \tfrac12   \gamma^{ab} F_{ab}^- \,\varepsilon_{ij} \epsilon^j + 2\,w
  A\,\eta_i\,,  \nonumber\\[.2ex]   
  \delta B_{ij} =&\,2\,\bar\epsilon_{(i} \Slash{D} \Psi_{j)} -2\,
  \bar\epsilon^k \Lambda_{(i} \,\varepsilon_{j)k} + 2(1-w)\,\bar\eta_{(i}
  \Psi_{j)} \,, \nonumber\\[.2ex] 
  \delta F_{ab}^- =&\,\tfrac12
  \varepsilon^{ij}\,\bar\epsilon_i\Slash{D}\gamma_{ab} \Psi_j+
  \tfrac12 \bar\epsilon^i\gamma_{ab}\Lambda_i
  -\tfrac12(1+w)\,\varepsilon^{ij} \bar\eta_i\gamma_{ab} \Psi_j \,,
  \nonumber\\[.2ex]   
  \delta \Lambda_i =&\,-\tfrac12\gamma^{ab}\Slash{D}F_{ab}^-
   \epsilon_i  -\Slash{D}B_{ij}\varepsilon^{jk} \epsilon_k +
  C\varepsilon_{ij}\,\epsilon^j 
  +\tfrac14\big(\Slash{D}A\,\gamma^{ab}T_{abij}
  +w\,A\,\Slash{D}\gamma^{ab} T_{abij}\big)\varepsilon^{jk}\epsilon_k
  \nonumber\\ 
  &\, -3\, \gamma_a\varepsilon^{jk}
  \epsilon_k\, \bar \chi_{[i} \gamma^a\Psi_{j]} -(1+w)\,B_{ij}
  \varepsilon^{jk}\,\eta_k + \tfrac12 (1-w)\,\gamma^{ab}\, F_{ab}^-
    \eta_i \,, \nonumber\\[.2ex]
    \delta C =&\,-2\,\varepsilon^{ij} \bar\epsilon_i\Slash{D}\Lambda_j
  -6\, \bar\epsilon_i\chi_j\;\varepsilon^{ik}
    \varepsilon^{jl} B_{kl}   \nonumber\\ 
  &\, -\tfrac14\varepsilon^{ij}\varepsilon^{kl} \big((w-1)
  \,\bar\epsilon_i \gamma^{ab} {\Slash{D}} T_{abjk}
    \Psi_l + \bar\epsilon_i\gamma^{ab}
    T_{abjk} \Slash{D} \Psi_l \big) + 2\,w \varepsilon^{ij}
    \bar\eta_i\Lambda_j \,. 
\end{align}

The independent fields of the Weyl multiplet transform
as follows,
\begin{eqnarray}
  \label{eq:weyl-multiplet}
    \delta e_\mu{}^a & =& \bar{\epsilon}^i \, \gamma^a \psi_{ \mu i} +
  \bar{\epsilon}_i \, \gamma^a \psi_{ \mu}{}^i \, , \nonumber\\ 
  \delta \psi_{\mu}{}^{i} & =& 2 \,\mathcal{D}_\mu \epsilon^i - \tfrac{1}{8}
  T_{ab}{}^{ij} \gamma^{ab}\gamma_\mu \epsilon_j - \gamma_\mu \eta^i
  \, \nonumber \\  
  \delta b_\mu & =& \tfrac{1}{2} \bar{\epsilon}^i \phi_{\mu i} -
  \tfrac{3}{4} \bar{\epsilon}^i \gamma_\mu \chi_i - \tfrac{1}{2}
  \bar{\eta}^i \psi_{\mu i} + \mbox{h.c.} + \Lambda^a_K e_{\mu a} \, ,
  \nonumber \\ 
  \delta A_{\mu} & =& \tfrac{1}{2} \mathrm{i} \bar{\epsilon}^i \phi_{\mu i} +
  \tfrac{3}{4} \mathrm{i} \bar{\epsilon}^i \gamma_\mu \, \chi_i +
  \tfrac{1}{2} \mathrm{i} 
  \bar{\eta}^i \psi_{\mu i} + \mbox{h.c.} \, , \nonumber\\  
  \delta \mathcal{V}_\mu{}^{i}{}_j &=& 2\, \bar{\epsilon}_j
  \phi_\mu{}^i - 3 
  \bar{\epsilon}_j \gamma_\mu \, \chi^i + 2 \bar{\eta}_j \, \psi_{\mu}{}^i
  - (\mbox{h.c. ; traceless}) \, , \nonumber \\   
  \delta T_{ab}{}^{ij} &=& 8 \,\bar{\epsilon}^{[i} R(Q)_{ab}{}^{j]} \,
  , \nonumber \\ 
  \delta \chi^i & =& - \tfrac{1}{12} \gamma^{ab} \, \Slash{D} T_{ab}{}^{ij}
  \, \epsilon_j + \tfrac{1}{6} R(\mathcal{V})_{\mu\nu}{}^i{}_j
  \gamma^{\mu\nu} \epsilon^j -
  \tfrac{1}{3} \mathrm{i} R_{\mu\nu}(A) \gamma^{\mu\nu} \epsilon^i + D
  \epsilon^i + 
  \tfrac{1}{12} \gamma_{ab} T^{ab ij} \eta_j \, , \nonumber \\ 
  \delta D & =& \bar{\epsilon}^i \,  \Slash{D} \chi_i +
  \bar{\epsilon}_i \,\Slash{D}\chi^i \, , 
\end{eqnarray}
where 
\be
  R(Q)_{\mu \nu}{}^i = & \, 2 \, \mathcal{D}_{[\mu} \psi_{\nu]}{}^i -
  \gamma_{[\mu}   \phi_{\nu]}{}^i - \tfrac{1}{8} \, T^{abij} \,
  \gamma_{ab} \, \gamma_{[\mu} \psi_{\nu]j} \ . 
\ee

Based on these two multiplets, one can write down a Lagrangian density 
for the chiral multiplet which is invariant under the superconformal transformations:
\begin{align}
  \label{eq:chiral-density}
  e^{-1}\mathcal{L} =&\, C - \varepsilon^{ij}\, \bar\psi_{\mu i} \gamma^\mu
  \Lambda_j-\tfrac18\bar \psi_{\mu i} T_{ab\,jk}\gamma^{ab}\gamma^\mu
  \Psi_l \,\varepsilon^{ij}\varepsilon^{kl}
   -\tfrac1{16}A( T_{ab\,ij} \varepsilon^{ij})^2 \nonumber\\
  &\, 
  -\tfrac12\bar\psi_{\mu i}\gamma^{\mu\nu}\psi_{\nu j}\,
  B_{kl}\,\,\varepsilon^{ik}\varepsilon^{jl} 
    + \varepsilon^{ij} \bar \psi_{\mu i}\psi_{\nu j}(F^{-\mu\nu}
    -\tfrac12 A\, T^{\mu\nu}{}_{kl}\,\varepsilon^{kl} )\nonumber\\
  &\,
  -\tfrac12 \varepsilon^{ij}\varepsilon^{kl} e^{-1}
  \varepsilon^{\mu\nu\rho\sigma} \bar\psi_{\mu i}\psi_{\nu j}
  (\bar\psi_{\rho k}\gamma_\sigma\Psi_{l} +\bar\psi_{\rho k}
  \psi_{\sigma j}\, A)\,.
\end{align}
This density is built such that the variation of the Lagrangian is equal to a total derivative
in spacetime. 

The Lagrangian for vector multiplets is based on first viewing the gauge invariant quantities 
of the vector multiplet as a reduced chiral multiplet with weight $w=1$. The components are:
\begin{align}
  \label{eq:vect-mult}
  A\vert_{\text{vector}}=&\,X\,,\nonumber\\
  \Psi_i\vert_{\text{vector}}=&\, \Omega_i\,,\nonumber\\
  B_{ij}\vert_{\text{vector}}=&\, Y_{ij}
  =\varepsilon_{ik}\varepsilon_{jl}Y^{kl}\,,\nonumber\\
  F_{ab}^-\vert_{\text{vector}}=&   \big(\delta_{ab}{}^{cd} -\tfrac12
    \varepsilon_{ab}{}^{cd}\big) e_c{}^\mu e_d{}^\nu \,\partial_{[\mu}
    A_{\nu]}\nonumber\\ 
    &\, 
  +\tfrac14\big[\bar{\psi}_{\rho}{}^i\gamma_{ab} \gamma^\rho\Omega^{j}
  + \bar{X}\,\bar{\psi}_\rho{}^i\gamma^{\rho\sigma}\gamma_{ab}
  \psi_\sigma{}^j
  - \bar{X}\, T_{ab}{}^{ij}\big]\varepsilon_{ij}  \,,\nonumber\\
  \Lambda_i\vert_{\text{vector}}
  =&\,-\varepsilon_{ij}\Slash{D}\Omega^j \,,\nonumber\\
  C\vert_{\text{vector}}= &\,-2\, \Box_\mathrm{c}  \bar X  -\tfrac14  F_{ab}^+\,
   T^{ab}{}_{ij} \varepsilon^{ij} - 3\,\bar\chi_i \Omega^i\,.
\end{align}
The transformations of the vector multiplet are:
\begin{align}
  \label{eq:variations-vect-mult}
  \delta X =&\, \bar{\epsilon}^i\Omega_i \,,\nonumber\\
  \delta\Omega_i =&\, 2 \Slash{D} X\epsilon_i
     +\half \varepsilon_{ij}  F_{\mu\nu}
   \gamma^{\mu\nu}\epsilon^j +Y_{ij} \epsilon^j
     +2X\eta_i\,,\nonumber\\
  \delta A_{\mu} = &\, \varepsilon^{ij} \bar{\epsilon}_i
  (\gamma_{\mu} \Omega_j+2\,\psi_{\mu j} X)
  + \varepsilon_{ij}
  \bar{\epsilon}^i (\gamma_{\mu} \Omega^{j} +2\,\psi_\mu{}^j
  \bar X)\,,\nonumber\\
\delta Y_{ij}  = &\, 2\, \bar{\epsilon}_{(i}
  \Slash{D}\Omega_{j)} + 2\, \varepsilon_{ik}
  \varepsilon_{jl}\, \bar{\epsilon}^{(k} \Slash{D}\Omega^{l)
  } \ . 
\end{align}

One then has to choose a meromorphic homogeneous function $F$ of weight $2$ and 
build the multiplet $\bf F(\bf X^{I})$ with lowest component $F(X^{I})$. 
The components of this is given in terms of the components of the vector multiplet as follows: 
\begin{align}
  \label{eq:chiral-mult-exp}
  A\vert_F =&\, F(A) \,,\nonumber\\
  \Psi_i\vert_F =&\, F(A)_I \,\Psi_i{}^I
  \,,\nonumber\\ 
  B_{ij}\vert_F =&\, F(A)_I\, B_{ij}{}^I -\tfrac12
  F(A)_{IJ} \,\bar \Psi_{(i}{}^I 
  \Psi_{j)}{}^J \,,\nonumber\\ 
  F_{ab}^-\vert_F =&\, F(A)_I \,F_{ab}^-{}^I -\tfrac18
  F(A)_{IJ}\, \varepsilon^{ij} \bar 
  \Psi_{i}{}^I \gamma_{ab} \Psi_{j}{}^J \,,\nonumber\\ 
  \Lambda_{i}\vert_F =&\, F(A)_I \,\Lambda_{i}{}^I
  -\tfrac12 
  F(A)_{IJ}\big[B_{ij}{}^I   \varepsilon^{jk} \Psi_{k}{}^J  
   +\tfrac12 F^{-}_{ab}{}^I\gamma^{ab} \Psi_{k}{}^J\big] \nonumber\\
   &\, 
   + \tfrac1{48}  F(A)_{IJK}\,\gamma^{ab} \Psi_i{}^I \,
   \varepsilon^{jk} \bar 
   \Psi_{j}{}^J \gamma_{ab}  \Psi_{k}{}^K \,,\nonumber\\ 
   C\vert_F =&\, F(A)_I\, C^I  -\tfrac14
   F(A)_{IJ}\big[ B_{ij}{}^I B_{kl}{}^J\, 
   \varepsilon^{ik} \varepsilon^{jl} 
   -2\, F^{-}_{ab}{}^I F^{-abJ} +4\,\varepsilon^{ik} \bar
   \Lambda_i{}^I \Psi_j{}^J\big]  \,,\nonumber\\ 
        &\,   +\tfrac14 F(A)_{IJK} \big[ \varepsilon^{ik}
        \varepsilon^{jl} 
        B_{ij}{}^I \Psi_{k}{}^J \Psi_{l}{}^K  -\tfrac12
        \varepsilon^{kl} \bar\Psi_{k}{}^I F^{-}_{ab}{}^J\gamma^{ab}
        \Psi_{l}{}^K\big] \nonumber\\ 
        &\,
        + \tfrac1{192} F(A)_{IJKL} \,\varepsilon^{ij}  \bar 
        \Psi_{i}{}^I \gamma_{ab} \Psi_{j}{}^J \,\varepsilon^{kl}  \bar
        \Psi_{k}{}^K \gamma_{ab} \Psi_{l}{}^L\,. 
\end{align}

\section{Boundary terms and supersymmetry of the renormalized action \label{Supersymmetry}}

In \S\ref{RenAction}, we conjectured the boundary action \eqref{Sbdry} 
\be
\CS_{\rm bdry} =  - 2 \pi r_{0} \left(  \frac{q_{I} \, e^{I}_{*}}{2} + i \, \big(F(X_{*}^{I}) -  \bar F(X_{*}^{I}) \big)  \right) \ . 
\ee
so that $\CS_{\rm ren}$ is finite. We also mentioned that 
this action is supersymmetric. In this appendix, we shall discuss the action $S_{\rm ren}$, and 
show that it is supersymmetric.

To motivate this, we note that we can rewrite $\CS_{\rm ren}$  as the sum of two pieces 
\bea\label{Srentwopieces}
\CS_{\rm ren} & = & \CS_{\rm bulk} + \CS_{\rm bdry} + i \frac{q}{2} \oint A \cr
& = & \left( \CS_{\rm bulk} +  \CS_{\rm bdry}^{1} \right) +  \left( \frac{i}{2}  q_I   \int_{0}^{2\pi} A^I_{\theta} \, d\theta +\CS^{2}_{\rm bdry} \right) \ , 
\eea
where we have split the boundary action \eqref{Sbdry} into a sum of two pieces:
\bea\label{Sbdrypieces}
\CS_{\rm bdry} & = & \CS_{\rm bdry}^{1} + \CS_{\rm bdry}^{2} \ , \\
\CS_{\rm bdry}^{1} & = & - \int_{0}^{2\pi}   i \, \Big[ F(X) - \bar{F (X)} \Big]_{\rm bdry} \, e^{\theta}_{\theta} \, d\theta \ ,  \\ 
\CS_{\rm bdry}^{2} & = & - \int_{0}^{2\pi}   \frac{q_I}{2} \,   \Big[ X^{I} + \bar X^{I} \Big]_{\rm bdry} \, e^{\theta}_{\theta} \, d\theta \ . 
\eea
Here,  $e^{\hat \theta} = \sinh{\eta_{0}}$ is the induced vielbein on the boundary. To verfiy
\eqref{Sbdrypieces}, we use the same algebra used in \eqref{Sbulk1}, namely, an expansion 
of the field $X^{I}$ into its fixed part $X^{I}_{*}$ and varying part which is $\CO(1/r_{0})$, followed by a Taylor 
expansion and the use of  attractor equations. 

With such a split of the action, the two pieces in \eqref{Srentwopieces} 
have a very natural interpretation as we discuss below. 
We will show further that each of them is finite and supersymmetric, implying the same for $\CS_{\rm ren}$. 

Recall that the bulk action \eqref{Sbulk} evaluated on the solution 
can be written as the difference of two pieces 
\be
\CS_{\rm bulk}  = 2 \pi i r_{0} \Big [F\big(X^{I}\big) 
- \bar{F\big(X^{I}\big)} \Big]_{\rm bdry}  
 - 2 \pi i\Big [F\big(X^{I}\big) - \bar{F\big(X^{I}\big)} \Big]_{\rm origin} \ . 
\ee
We see that $\CS_{\rm bulk}$ + $\CS^{1}_{\rm bdry}$ is manifestly finite. 
Thus, $\CS^{1}_{\rm bdry}$ has the natural interpretation of a canonical boundary term 
which cancels the boundary part of the bulk action, so that any variation of 
$\CS_{\rm bulk}$ + $\CS^{1}_{\rm bdry}$ will be finite and not contain boundary terms.

The second piece of the boundary action combines with the Wilson line to give the operator 
\be\label{Swilsonsugra}
\exp \big[ - \frac{i}{2}  q_I   \int_{0}^{2\pi} A^I_{\theta} \, d\theta - \CS^{2}_{\rm bdry} \big]
 =   \exp \big[  -\frac{i}{2} \, q_I \int_{0}^{2\pi}  \left(A^I_{\theta} + i e^{\theta}_{\theta} \, (X^{I} + \bar X^{I}) \right)  
d\theta   \big] 
\ee
This operator has the natural interpretation as the supersymmetric Wilson line of 
gauge theory \cite{Maldacena:1998im, Rey:1998ik}. Recalling the boundary behavior of the fields 
\bea
- \frac{i}{2}  q_I   \int_{0}^{2\pi} A^I_{\theta} \, d\theta & = &  - \pi \, q_I \, e_{*}^I \, r_{0}(1 + \CO(1/r_{0})) \ ,  \\
- \frac{i}{2} \, q_I \int_{0}^{2\pi}  i e^{\theta}_{\theta} \, (X^{I} + \bar X^{I})  d\theta & = & 
 \pi  \, q_{I} \, r_{0}\Big( X^I_{*} + \bar X^I_{*} + \CO(1/r_{0}) \Big)   \ ,  \\
& = & \pi \, q_I \, e_{*}^I \, r_{0}(1 + \CO(1/r_{0})) \ , 
\eea
it is easy to see that this operator is manifestly finite. 

Evaluated on the solutions $A^{I}_{\theta} = - i e^{I}_{*}(r_{0}-1)$, $X^{I} = X^{I}_{*} + \frac{C^I}{r_{0}}$, $\bar X^{I} = \bar X^{I}_{*} + \frac{C^I}{r_{0}} $ that we consider in \S\ref{Solution}, we see that the two pieces of the 
renormalized action \eqref{Srentwopieces} above give the two pieces of the final renormalized action 
\eqref{Srenfinal} which we found in \S\ref{RenAction}, as indeed should happen.

In the rest of the appendix, we shall sketch the proof of supersymmetry of these two operators.
The supersymmetry of the operator  \eqref{Swilsonsugra} above follows from the transformation 
rules of $X^{I}$ and $A_{\mu}^{I}$ of the vector multiplet \eqref{eq:variations-vect-mult}. 
We use the fact that the Killing spinors obey 
\be\label{killingsprel}
\zeta^{i} = \varepsilon_{ij} \gamma^{0} \zeta^{j} \ . 
\ee
The extra term in the variation of the vector field which is proportional to the gravitino 
is cancelled by the variation of the vielbein in the definition of the super Wilson line. 
This is the new ingredient in the super Wilson line of a gravitational theory compared to that of gauge 
theory. 

Now we come to the supersymmetry of the combination $\CS_{\rm bulk} + \CS^{1}_{\rm bdry}$.
The statement that $\CS_{\rm bulk}$ is supersymmetric \cite{deWit:1979ug, deWit:1984px, deWit:1980tn} 
really means that the 
variation of $\CS_{\rm bulk}$ is a boundary term which can be ignored in certain circumstances. 
In our situation, there is a non-trivial boundary, and therefore what we need to check is that the 
variation of the bulk Lagrangian is indeed equal to the derivative of the boundary Lagrangian. 

To investigate this, we need to understand the structure of the Lagrangian built using the 
chiral superfield \cite{deRoo:1980mm}. 
In the case of rigid supersymmetry, the variation of the top component of the chiral superfield 
is a total derivative in spacetime, and therefore the top component (picked by a chiral superspace integral) 
is an invariant Lagrangian. 
For chiral superfields coupled to superconformal gravity, the transformation rules undergo 
a modification and the derivatives become covariant derivatives, and there are additional 
terms in the variation \eqref{eq:conformal-chiral} of the top component $C$. 
The invariant  Lagrangian density  \eqref{eq:chiral-density} contains 
new terms whose variation cancel the additional non-derivative terms present in $\delta C$. 

The net result of this procedure is that the variation of the invariant 
Lagrangian is equal to the total derivative terms that are present in the variation of the top component 
$C$ of the chiral multiplet, {\it i.e.} essentially one can drop the extra terms which arise due to 
the covariantization. 
As an example,  the term proportional to the auxiliary field $B_{ij}$ in $\delta C$ contains $\chi_{i}$
which is an auxiliary field of the superconformal multiplet constrained to be proportional 
to $R(Q)^i$. This term is cancelled by the term proportional to $B_{ij}$ in the higher corrections 
to the Lagrangian density \eqref{eq:chiral-density} after solving for the auxiliary field $\chi$ in terms of 
the gravitini. 

%

Looking at the $Q$ variation \eqref{eq:conformal-chiral} of a chiral multiplet of weight $w=2$, 
we see that the variation of $C$ contains two total derivative pieces 
\be\label{totalder1}
-2 \varepsilon^{ij} \slash{\p} (\bar\epsilon_i \, \Lambda_j ) \ , 
\ee 
and 
\be\label{totalder2}
 -\tfrac14\varepsilon^{ij}\varepsilon^{k \ell} \big( ({\slash{\p}}  \bar\epsilon_i \, \gamma^{ab} \, T_{abjk})
\Psi_l + \gamma^{ab} \, T_{abjk} \, \slash{\p} (\bar\epsilon_i \, \Psi_\ell )\big)  
=  -\tfrac14\varepsilon^{ij}\varepsilon^{k \ell} \, {\slash{\p}} \,
( \bar\epsilon_i \, \gamma^{ab} \, T_{abjk} \Psi_\ell \big)  \ . 
\ee  

In our problem where we have a bunch of vector multiplets, the way to build a Lagrangian 
is by using the homogeneous function $F(X^{I})$. One 
first builds a chiral multiplet $\bf F (\bf X^{I})$ whose bottom component is $F(X^{I})$,
and then uses the invariant Lagrangian described above for this chiral multiplet. 
The variation of our Lagrangian is therefore equal to the total derivative terms that appear 
in the variation of the top component of the chiral multiplet $\bf F (\bf X^{I})$.  
Looking at the form of the components of this superfield \eqref{eq:chiral-mult-exp}, and then 
substituting for the components of the reduced chiral multiplet \eqref{eq:vect-mult}, 
we find that the first type of total derivative term from integration of \eqref{totalder1} is 
\bea
&-& 2 \varepsilon^{ji} \int_{\rm bulk} \slash{\p} (\bar\epsilon_j \Lambda_i )\vert_{F}\cr
 &=& 
-2 \varepsilon^{ji}  \int_{\rm bulk}  \slash{\p} \bigg( - \bar\epsilon_j  F(X)_I \, \varepsilon_{ik}\Slash{D}\Omega^{kI}
  -\tfrac12 
  \bar\epsilon_j F(X)_{IJ}\big[B_{ij}{}^I   \varepsilon^{jk} \Omega_{k}{}^J  
   + \tfrac12 \bar\epsilon_j F^{-}_{ab}{}^I\gamma^{ab} \Omega_{k}{}^J\big]     \nonumber\\
   && + \tfrac1{48}  \bar\epsilon_j \, F(X)_{IJK}\,\gamma^{ab} \, \Omega_i{}^I \,
   \varepsilon^{k\ell} \, \bar 
   \Omega_{k}{}^J  \gamma_{ab} \, \Omega_{\ell}{}^K \bigg) \ . 
\eea

We are interested in the bosonic part of the boundary counterterm Lagrangian. 
The third term on the right hand side contains three fermions and so cannot 
appear from the variation of a pure bosonic term, so we shall ignore that term here.
The second term on the RHS proportional to $F_{IJ}$ is equal to the variation of 
$F_{IJ} \Omega^{I} \Omega^{J}$ minus a total derivative term on the boundary. We 
can therefore drop this term since it is fermionic. Using the variation 
$\delta X^{I} = \bar{\epsilon}^i\Omega_i^{I}$, the first term on the RHS is 
proportional to the variation of the derivative of $F_{I}$, which integrates to zero 
 on the closed boundary, and therefore does not produce any boundary counterterms.

This leaves us with the second term \eqref{totalder2} which gives rise to a boundary term 
\be\label{deltaCfin}
-\tfrac14\varepsilon^{ij}\varepsilon^{k \ell} \int_{\rm bulk}  \slash{\p} \, 
 \big( \bar\epsilon_i \, \gamma^{ab} \, T_{abjk} \, \Psi_\ell \vert_{F} \big)  
= -\tfrac14\varepsilon^{ij}\varepsilon^{k \ell} \int_{\rm bulk}  \slash{\p} \, 
\big(  \bar\epsilon_i \, \gamma^{ab} \, T_{abjk} \, F(X)_I \,\Omega_\ell{}^I \big) \ . 
\ee 
Now, the variation of $T_{abjk}$   \eqref{eq:weyl-multiplet} is proportional to the 
curvature $R(Q)_{ab}$ which integrates to zero on the boundary. Therefore, $T_{abjk}$ can be 
treated as a constant on the boundary for the purpose of supersymmetry 
variations. Plugging in the attractor value for $T_{abjk}$, and 
using $\delta X^{I} = \bar{\epsilon}^i\Omega_i^{I}$ again, and the Killing spinor relation \eqref{killingsprel}, 
we see that the remaining term \eqref{deltaCfin}
is equal and opposite to the variation of the boundary term $\CS^{1}_{\rm bdry}$, thus showing that 
the supersymmetry variation of $\CS_{\rm bulk} + \CS^{1}_{\rm bdry}$ vanishes.

\bibliographystyle{JHEP}
\bibliography{final}

\end{document}